\documentclass[twocolumn]{aastex631}

\usepackage{amsmath}
\usepackage{dsfont}

\newcommand{\dd}{{\rm d}}
\newcommand{\ee}{{\rm e}}

\begin{document}

\title{Radial modes of pressure bumps and dips in astrophysical discs}

\author[0000-0001-6136-4164]{Armand Leclerc}
\affiliation{Institute of Science and Technology Austria (ISTA), Am Campus 1, 3400 Klosterneuburg, Austria}

\author[0000-0002-3116-3166]{Guillaume Laibe}
\affiliation{ENS de Lyon, CRAL UMR5574, Universite Claude Bernard Lyon 1, CNRS, Lyon, F-69007, France}

\author{Elliot Lynch}
\affiliation{ENS de Lyon, CRAL UMR5574, Universite Claude Bernard Lyon 1, CNRS, Lyon, F-69007, France}

\author{Nicolas Perez}
\affiliation{Department of Geophysics, Porter school of the Environment and Earth Sciences, Tel Aviv University,69978 Tel Aviv, Israel}

\begin{abstract}
This study investigates the signatures of pressure extrema on global oscillations in discs. To this end, we use the framework of wave topology to establish a generalised local dispersion relation that includes pressure gradients. We highlight the influence of a previously unrecognized epicyclic–acoustic frequency and derive an analytical criterion for the existence of a branch of modes transiting between the inertial and the pressure bands. We find that pressure extrema consist of wave guides in which such \textit{topological modes} propagate. The fundamental mode trapped at a pressure bump can propagate at all frequencies, allowing it to resonate with any temporal forcing, while the mode associated with a pressure gap propagates at a fixed frequency, propagates with arbitrary vertical phase velocity. These specific features make them attractive candidates for future discoseismology.
 
\end{abstract}


\section{Introduction} 
\label{sec:intro}
Solids in protoplanetary discs drift towards regions of higher pressure \citep{Safronov1972}. As such, pressure maxima -- often called pressure bumps -- are privileged places to collect solids and foster planet formation (e.g. \citealt{Bae2023,Lesur2023,Dra2023}), with locating and characterising such features of great interest to the planet formation community. Only dust can be imaged directly, while the gas pressure profile has recently been quantitatively constrained from inverting the mean rotational profile, using high-precision kinematic data of CO emission lines obtained with ALMA (e.g. \citealt{Pinte2023}). Departures from circular Keplerian motion can arise from pressure gradients, disc eccentricity or inclination, secondary flows, or additional physical processes such as self-gravity, dust dynamics, or magnetic fields (e.g. \citealt{Teague2025}). These processes and the presence of protoplanets shape the pressure profile of the gas into a structured disc, with pressure bumps and dips. Probing this flow and these processes by a seismic analysis could be achieved with spatially or time-resolved kinematic data: this is the very goal of \textit{discoseismology}, essentially introduced so far to study acretion discs around black holes \citep{Solheim1998,Tsang2009,Tsang2013,OR2020,Dewberry2020a,Dewberry2020b,Kato2024}. This approach aims at constraining the structure of the disc not via the mean flow, but through the study of linear perturbations about its equilibrium.

Linear waves and instabilities have a well-established history as mechanisms for forming structures in discs \citep{toomre1964,goldreich1965a,lyndenBell1967,papaloizou1985}, for instance the spiral arms of galaxies are understood to be density waves \citep{lin1966,lynden1972}. External perturbers can excite both inertial and acoustic waves at Lindblad resonances, where the orbital period of the mean flow is commensurable with the perturbation's period, which launches waves in the disc \citep{lindblad1948,goldreich1978,goldreich1979,lubow1998}. This occurs in Saturn's rings, where ring waves in resonance with oscillations of the planet have been observed by Cassini which provided the first and only seismic data of Saturn \citep{marley1991nonradial,hedman2013,fuller2014saturn}.

While global modes of isothermal and polytropic discs have been widely investigated \citep{kato1980,kato1983,papaloizou1985,blaes1985,blaes2006,blaes2007}, the analysis becomes simpler when focusing solely on small-scale perturbations within the shearing-box framework, where the fluid evolution is described in a small cartesian box rotating with the flow \citep{hill1878,goldreich1965}. This has led to important advances in our understanding of wave propagation, instabilities, turbulence, angular-momentum transport and planetesimal formation in accretion and protoplanetary discs (e.g.  \citealt{kato1978,goldreich1980,balbus1991,lubow1993,goodman1993,korycansky1995,ogilvie1998,balbus1998,ogilvie1999,youdin2005,johansen2007,nelson2013}).

On the other hand, as in the case in stellar seismology (e.g. \citealt{Aerts2010}), key targets for discoseismology are the large-scale global modes, since they probe the overall structure of the disc and are the easiest to detect. Among these, \textit{topological modes} are especially attractive, since they can be studied and constrained without the need for short-wavelength or local techniques. These modes stem from a topological property of the differential operator that governs their evolution. While such modes have been investigated for decades in condensed-matter physics, their study in hydrodynamics has emerged only recently \citep{delplace2017,perrot2019,perez2022,parker2020,qin2023,Perez2025}. Since topological analysis has received little attention in astrophysical fluids to date, characterization of topological modes may enable new methods of probing the structure of interesting objects. The example of topological waves in stars is to this respect particularly instructive. \citet{leclerc2022} demonstrated that the cancellation of a particular characteristic frequency in stars necessarily gives rise to a topological mode, explaining why $f$-branch modes with small harmonic degrees are not confined to the stellar surface. These modes were subsequently shown to remain robust in the presence of convective regions, radiative dissipation, and are excited by convective motions \citep{Leclerc2024a,LL2025}. They have also been observed to hybridize with $g$-modes, providing the most reliable method to date for inferring the rotation rate of the solar core \citep{LL2025}.  
Stellar rotation was also found to induce topological modes in seismic spectra \citep{Leclerc2024c}.

Building on these developments performed in the stellar context, we aim to address the following questions: do topological modes exist in discs? If so, how are they related to the pressure profile? We address these questions here, and highlight the special properties these modes have, suggesting their interest for protoplanetary discs seismology and dynamics. In Sect.~\ref{sec:microlocal}, we start by examining the local propagation of waves, by performing a so-called microlocal analysis of the linear wave operator of a simple model of a disc to obtain the local dispersion relation as well as the local polarization relations between the wave fields \citep{keppeler2004,hall2013,onuki2020,vidal2024}. This is the first time that this method is used in the context of astrophysical discs. In Sect.~\ref{sec:global_modes}, we exhibit the topological properties of the problem, and predict the existence of topological modes. We then focus on the role of the presence of extrema in the pressure profile in Sect.~\ref{sec:extrema}, analysing how they are reflected in the spectrum of large-scale modes. Further extensions of this study are discussed in Sect.~\ref{sec:discussion}.

\section{Local waves without shearing box}
\label{sec:microlocal}

\subsection{Epicyclic-acoustic frequency}

We consider an unstratified disc made of a non-self gravitating ideal gas around a star of mass $M$. The equations of motion for the velocity $\mathbf{v}$, pressure $p$ and density $\rho$ are
\begin{eqnarray}
    \partial_t \mathbf{v} + \left(\mathbf{v}\cdot\nabla \right)\mathbf{v} &=& -\frac{1}{\rho}\nabla p -\frac{\mathcal{G}M}{r^2}\boldsymbol{e}_r,\label{eq:fulldyn1}\\
    \partial_t \rho + \nabla \cdot \left( \rho \mathbf{v}\right) &=& 0,\label{eq:fulldyn2}
\end{eqnarray}
along with an appropriate energy equation.
We restrict our analysis to the case of a circular disc. In cylindrical coordinates $(r,\theta,z)$, the equilibrium profile is parametrized by $\rho = \rho_0(r)$, $p = p_0(r)$, $v = (0,r\Omega(r),0)$. Along the radial direction, 
\begin{equation}
    r\Omega^2 = \frac{1}{{\rho_0}}\frac{\dd p_0}{\dd r} +\frac{\mathcal{G}M}{r^2}.
\end{equation}
\label{subsec:globalWaves}
We linearly expand the quantities as $X = X_0 + X^\prime$ in Eqs.\eqref{eq:fulldyn1}-\eqref{eq:fulldyn2}, assuming adiabatic axisymmetric perturbations of the gas ($\delta p=c_\mathrm{s}^2\delta\rho$). One obtains  \citep{latter2017local}
\begin{eqnarray}
    \partial_t v_r^\prime &=& 2\Omega v_\theta^\prime - \frac{1}{\rho_0}\partial_rp^\prime + \frac{p^\prime}{c_\mathrm{s}^2 \rho_0^2}\frac{\dd p_0}{\dd r},\\
    \partial_t v_\theta^\prime &=& -\left(2\Omega+r\frac{\mathrm{d}\Omega}{\mathrm{d}r}\right) v_r^\prime,\\
    \partial_t v_z^\prime &=& -\frac{1}{\rho_0} \partial_z p^\prime,\\
    \partial_t p^\prime &=& - p_0\left(\frac{1}{r}\partial_r (r v_r^\prime) + \partial_z{v_z^\prime}\right) - \frac{\dd p_0}{\dd r} v_r^\prime.
\end{eqnarray}
We apply the change of variables
\begin{eqnarray}
(v_r^\prime,v_\theta^\prime,v_z^\prime) &\mapsto& \boldsymbol{u} \equiv \sqrt{r\rho_0}\left(v_r^\prime,\sqrt{\frac{2}{2-s}}v_\theta^\prime,v_z^\prime\right), \label{eq:variableschange}\\
    p^\prime &\mapsto& h \equiv \frac{\sqrt{r}}{\sqrt{\rho_0}c_\mathrm{s}}p^\prime,
\end{eqnarray}
where $s \equiv \frac{\dd\ln\Omega}{\dd\ln r}$. It reveals a symmetrized set of equations, which simplifies the subsequent choice of inner product introduced for the topological analysis (\citealt{leclerc2022}, see Sect.~\ref{subsec:localWaves}). Performing a Fourier transform $\ee^{i\omega t - i k_z z}$ yields a wave equation for ${X^\top \equiv \begin{pmatrix}u_z & u_r & u_\theta & h \end{pmatrix}}$ of the form
\begin{eqnarray}
    &&\mathcal{H}X = \omega X ,\label{eq:eigenvalGlobalwaves}\\
    &&\mathcal{H} = \label{eq:globalWaves}\\
    &&\begin{pmatrix}
        0 & 0 & 0 & c_\mathrm{s}k_z \\
        0 & 0 & -i\kappa & \;\;ic_\mathrm{s}\partial_r + \frac{i}{2}\frac{\dd c_\mathrm{s}}{\dd r} - i S \\
        0 & i\kappa & 0 & 0 \\
        c_\mathrm{s}k_z &\;\; ic_\mathrm{s}\partial_r + \frac{i}{2}\frac{\dd c_\mathrm{s}}{\dd r} + i S & 0 & 0 
    \end{pmatrix}\nonumber,
\label{eq:schro}    
\end{eqnarray}
where $\kappa \equiv \sqrt{2(2-s)}\Omega$ is the epicyclic frequency and 
\begin{equation}
S \equiv \frac{c_\mathrm{s}}{2}\left(\frac{\dd\ln p_0}{\dd r} + \frac{\dd\ln c_\mathrm{s}}{\dd r} + \frac{1}{r}\right)
\label{eq:S}
\end{equation}
denotes the \textit{epicyclic-acoustic frequency} of the disc, acting as a cutoff frequency as we will show in the local dispersion relation. To the best of our knowledge, this is the first explicit identification of the frequency that governs momentum exchange between the inertial and acoustic bands in discs. Its expression follows directly from the rescaled symmetric Hermitian formulation of the wave equations. $S$ depends on the radial inhomogeneity and the curvature of the disc. Terms associated to Eq.~\eqref{eq:schro} break the symmetry $(r,u_r,u_\theta)\to -(r,u_r,u_\theta)$, which is only restored when $S=0$. $S$ plays the role, in discs, of the buoyant–acoustic frequency identified in stars by \citep{leclerc2022}. For a stratified disc, this expression can be conveniently rewritten
\begin{equation}
S = \frac{1}{2}\left[ \frac{\mathrm{d} \ln p_{0} }{\mathrm{d} \ln r} + \frac{\mathrm{d} \ln c_{\rm s} }{\mathrm{d} \ln r} + 1 \right] \frac{H_P}{r} \Omega ,
\label{eq:Salt}
\end{equation}
where $H_P \equiv c_{\rm s} \Omega^{-1}$ is the pressure scale height of the disc.

We impose impenetrable boundaries, such that $u_r(r_0)=u_r(r_1)=0$. For real $\kappa$ (a Rayleigh stable disc), the operator $\mathcal{H}$ is self-adjoint with respect to the canonical inner product $\langle X,Y\rangle = \int \dd r\; X^{\dagger} Y$. The frequencies $\omega$ are then guaranteed to be real: the perturbations considered here are linearly stable. The self-adjointness of $\mathcal{H}$ gives the following energetic constant of motion
\begin{eqnarray}
    I &=& \frac{1}{2}\langle X,X\rangle,\\
    &=& \int \dd r \; \frac{1}{2}\left(\vert \boldsymbol{u}\vert ^2 + \vert h\vert ^2\right) ,\label{eq:constantOfMotion}\\
    &=& \int \dd r \;r\left[ \frac{1}{2}\rho\left(\vert v^{\prime}_r\vert ^{ 2} + \frac{2}{2-s}\vert v^{\prime}_\theta\vert ^{ 2} + \vert v^{\prime}_z\vert ^{2}\right) + \frac{1}{2\rho c_\mathrm{s}^2}\vert p^{\prime}\vert ^{2}\right] ,\nonumber
\end{eqnarray}
(see Appendices for a proof). The solutions to Eqs.\eqref{eq:eigenvalGlobalwaves}-\eqref{eq:globalWaves}, together with these boundary conditions, are the axisymmetric oscillation modes of the disc, i.e. the standing waves in the radial direction.

\subsection{Local wave equation}
\label{subsec:localWaves}
Microlocal analysis is a representation of differential operators on a phase space using the Wigner transform \citep{keppeler2004,hall2013}. This transform maps (pseudo-)differential operators to {\it symbols}, i.e functions of the phase space. Three useful symbols used here are
\begin{align}
    \mathrm{Symb}[c_\mathrm{s}(r)]\quad\quad &= c_\mathrm{s},\\
    \mathrm{Symb}[i\partial_r] \quad\quad&= k_r,\\
    \mathrm{Symb}[ic_\mathrm{s}\partial_r + \frac{i}{2}\frac{\dd c_\mathrm{s}}{\dd r}] &= c_\mathrm{s}k_r.
\end{align}
It gives a appropriate way to define a local wavenumber $k_r$ for a wave \citep{onuki2020}.\\
The operator $\mathcal{H}$ is then mapped to its symbol $H$
\begin{equation}
    H = \begin{pmatrix}
        0 & 0 & 0 & c_\mathrm{s}k_z \\
        0 & 0 & -i\kappa & c_\mathrm{s}k_r - i S \\
        0 & i\kappa & 0 & 0 \\
        c_\mathrm{s}k_z & c_\mathrm{s}k_r + i S & 0 & 0 
    \end{pmatrix}.
    \label{eq:symbolH}
\end{equation}
The Wigner transform has a number of useful properties, among which is the conservation of Hermiticity: 
$\mathcal{H}$ is a self-adjoint differential operator if and only if $H$ is a Hermitian matrix function (one can find this property in \citep{hall2013}, page 266). It is then guaranteed that $H$ gives the local dispersion relation of the waves without introducing spurious instabilities, as it has been shown to happen when including pressure gradients in local studies (e.g. see discussions in \citealt{lin2015,latter2017local}). Indeed, one has $H^{\dagger} = H$. The local wave equation is
\begin{equation}
    HX = \omega X,\label{eq:localWaveEq}
\end{equation}
and the local dispersion relation is simply obtained by $\det\left(H - \omega\mathds{1}\right)=0$, which yields
\begin{equation}
    c_\mathrm{s}^2k_r^2 = \frac{(c_\mathrm{s}^2k_z^2-\omega^2)(\kappa^2-\omega^2)}{\omega^2} - S^2. \label{eq:localDisp}
\end{equation}
Equation~\eqref{eq:localDisp} is a quadratic polynomial in $\omega^2$, whose smaller root corresponds to the inertial band (or $r$-band) and the larger to the pressure band (or $p$-band). For inertial waves, one has ${0 \leq \omega^2 \leq \kappa^2}$, while for pressure waves, ${\omega^2 \geq \kappa^2+S^2}$. Thus, a nonzero value of $S$ opens a frequency gap between the bands, and no wave with $\omega^2 \in (\kappa^2,\kappa^2+S^2)$ can propagate radially in the disc. The dispersion relation derived by other local expansions, which neglect both the curvature term and the pressure gradient, is recovered for $S=0$. In that special case, there is a degeneracy between acoustic and inertial waves at $\omega = \kappa$ for $(k_r,k_z)=(0,\kappa/c_\mathrm{s})$, but it is lifted by any nonzero value of $S$, i.e any nonzero pressure gradient or by the curvature term.

\section{Topology and global modes}
\label{sec:global_modes}

\subsection{Chern numbers}
\label{subsec:chern}
The symbol $H$ obtained from the microlocal analysis formally coincides with that derived for stellar oscillations in \citet{leclerc2022}, who demonstrated that these degeneracies in the local dispersion relation are linked to topological charges known as Chern numbers \citep{volovik2003,delplace2022}. According to the principle called index theorem, this implies the presence of a spectral flow in the dispersion relation of the global modes $\{\omega_n\}_{n \in \mathds{N}}$ \citep{delplace2017,delplace2022,faure2023}. A spectral flow is the fact that a finite number of modes branches transit from one band to the other as the parameter $k_z$ increases, in that case between the band of $r$-modes and the band of $p$-modes. This number of branches precisely equals a topological index of the symbol called the first Chern number.

\begin{figure}
    \centering
    \includegraphics[width=\columnwidth]{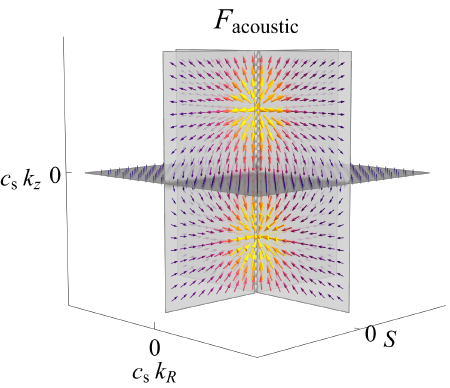}
    \caption{The Berry curvature $\boldsymbol{F}$ of the acoustic wave is singular at $(c_\mathrm{s}k_r,S,c_\mathrm{s}k_z)=(0,0,\pm\kappa)$. This obstruction is a topological constraint, characterized by the two charges $\mathcal{C}=\pm 1$. Length and brightness of the arrows indicate the norm of $\boldsymbol{F}$.}
    \label{fig:berryCurv}
\end{figure}

Following \citet{leclerc2022}, we define and compute the Chern numbers for the wavebands in the disc. For a given eigenvector $X$ of a given band of $H$ (either the $r$- or the $p$- band), the Berry curvature is the vector field in the parameter space $(k_r,S,k_z)$ expressed as
\begin{equation}
    \boldsymbol{F} = i \boldsymbol{\nabla} \times \left( X\cdot \boldsymbol{\nabla}X\right),
\end{equation}
where $ \boldsymbol{\nabla} = (\partial_{k_r},\partial_S,\partial_{k_z})$. This convenient expression only holds in a 3D parameter space, although more general definitions exist \citep{delplace2022}. $\boldsymbol{F}$ is singular when frequencies degenerate, which occurs for ${k_r=S=0}$ and $c_\mathrm{s}k_z=\pm\kappa$. Fig.~\ref{fig:berryCurv} displays $\boldsymbol{F}$ for the acoustic wave, revealing the two singularities associated with the degeneracies.\\
The Chern number characterizes these singularities. It is an integer defined as 
\begin{equation}
    \mathcal{C} = \frac{1}{2\pi}\oint_\Sigma \boldsymbol{F}\cdot \dd \boldsymbol{\Sigma},
\end{equation}
where $\Sigma$ is any closed surface enclosing one singularity. Numerical computations show that the two degeneracies at $k_r=S=0$ and ${c_\mathrm{s}k_z=\pm\kappa}$ have Chern numbers 
\begin{equation}
\mathcal{C}=\pm 1.
\end{equation}
These values are expected from the formal analogy with the problem of \citet{leclerc2022}.\\
The Chern number is computed in the local analysis; however, it has direct consequences on the global waves by imposing a spectral flow, the existence of modes whose frequencies approach both wavebands (inertial and acoustic) for different limits of $k_z$. The non-zero Chern numbers calculated by \cite{leclerc2022} impose the propagation of a Lamb-like wave in non-rotating stars crossing the frequency gap between acoustic and internal gravity waves \citep{perrot2019,leclerc2022}. 
We then expect to have one global wave branch with such spectral behavior in the spectrum of a disc.

\subsection{Location of topological modes}
\label{sec:loc}

The spatial location of topological modes corresponding to the spectral flow branch is determined by whether a radius exists where 
\begin{equation}
S = 0 ,
\end{equation}
as it is the point where the topological degeneracy occurs in Fig.~\ref{fig:berryCurv}. If such a topological interface is present in the disc, the energy of the modes are concentrated in a volume centered around this interface over a typical radial length $\mathcal{L}$ given by
\begin{equation}
\mathcal{L} \equiv \sqrt{c_{\mathrm{s}} /\left|\frac{\mathrm{d} S}{\mathrm{~d} r}\right|_{S=0}}.
\label{eq:trapping}
\end{equation}
The steeper $S$ is near the topological interface, the more tightly confined the topological modes are. If $S$ has a constant sign throughout the object, the modes are located at one of the boundaries: at the inner edge if $S < 0$ and at the outer edge if $S > 0$ (See the discussion of \citep{iga2001}, page 483 for a similar result in an atmospheric context). This behavior is a manifestation of the general bulk-boundary correspondence known in condensed matter physics: the topological properties of a material, calculated in the bulk of the domain, determine the existence and nature of some modes that appear at its boundaries \citep{delplace2022}.

\subsection{Monotonic disc profiles}
\label{subsec:monotonic}
We first consider a disc whose density, temperature and rotation rate are parametrized by typical power-laws \citep{armitage2011}. More precisely, the temperature scales as $T(r) \propto r^{-q}$, the surface density as $\Sigma(r) \propto r^{-p}$ and the rotation rate as $\Omega(r) \propto r^{-s}$. Hence, volume density scales as $\rho \propto \frac{\Omega}{c_\mathrm{s}}\Sigma \propto r^{-p-s+\frac{q}{2}}$ and
\begin{equation}
S = \frac{1}{2}\left[1 - p - q - s \right] \frac{H}{r} \Omega .
\label{eq:Smono}
\end{equation}
For physically realistic profiles, $1 - p - q - s<0$ and thus $S < 0$ throughout the disc. As discussed in Sect.~\ref{sec:loc}, a topological mode is expected to be located at the inner edge of the disc.\\ 

\begin{figure}
    \centering
    \includegraphics[width=\columnwidth]{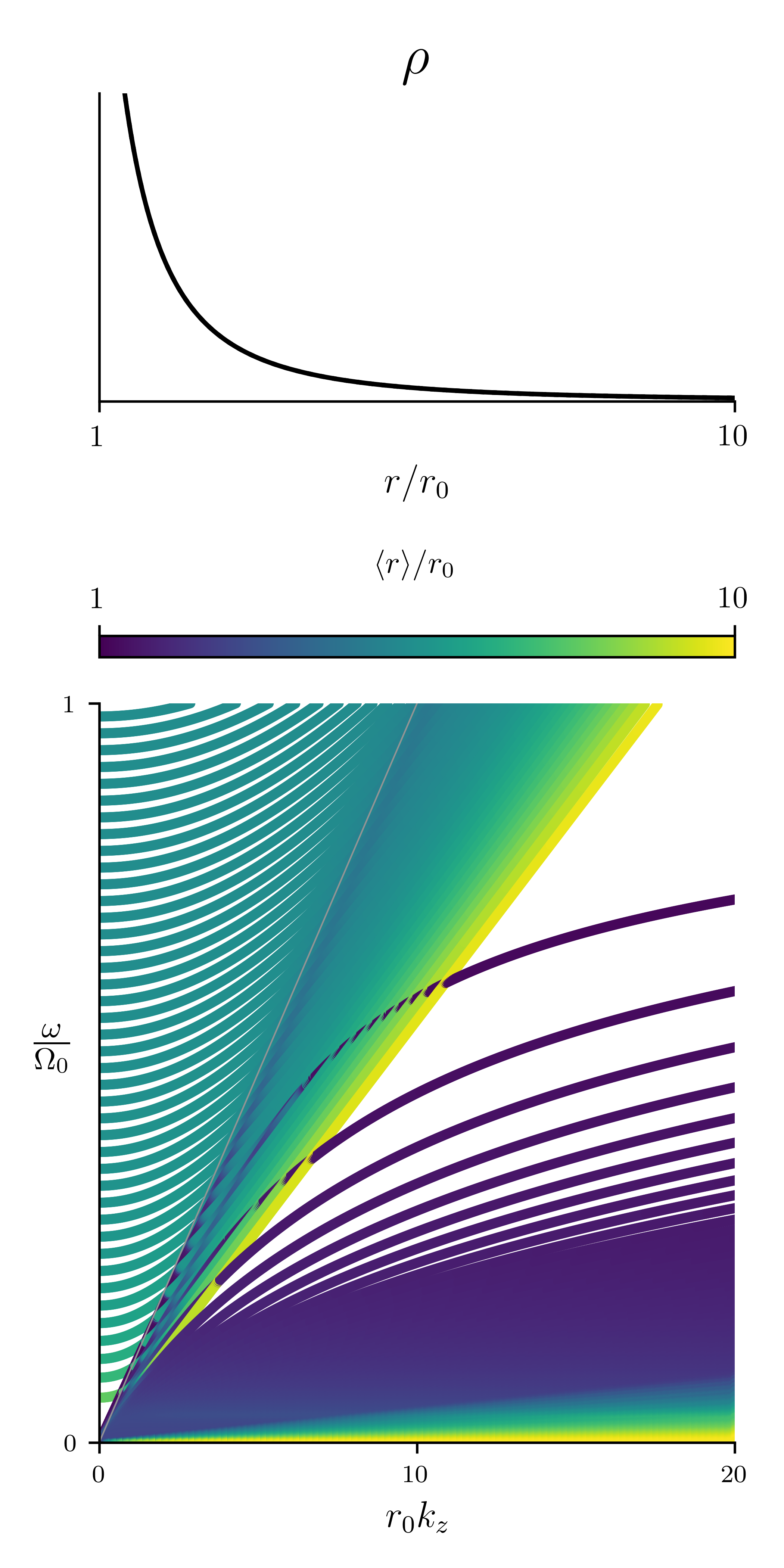}
    \caption{Global modes of a monotonic disc. The one topological branch transits from inertial modes (blue) at low $k_z$ to acoustic modes (yellow) at high $k_z$. It matches the dispersion relation $\omega=c_\mathrm{s}(r_0)k_z$ (grey thin line).}
    \label{fig:monotonic}
\end{figure}

We verify this prediction by solving Eqs.~\eqref{eq:eigenvalGlobalwaves}-\eqref{eq:globalWaves} numerically using the \texttt{EVP} class of the \textsc{Dedalus} python package \citep{burns2020,oishi2021}. The gas is contained between two radii $r_0$ and $r_1 = 10 r_0$. Units of lengths and time are $r_0$ and $\Omega(r_0)^{-1}$ and the dimensionless parameters are $p=1$, $q=\frac{1}{2}$, $s=\frac{3}{2}$ and $\frac{c_\mathrm{s}}{R\Omega}\big|_{r_0}=0.1$. Figure~\ref{fig:monotonic} shows the density profile of the disc and the frequencies of the global waves for a range of vertical wavenumbers $k_z$. The average energy-weighted position ${\langle r\rangle \equiv \langle X,rX\rangle/\langle X,X\rangle}$ for a mode $X(r)$ is provided via the color bar. Low frequency $p$-modes are located mostly in the outer parts of the disc, where the sound speed is the lowest. High frequency $r$-modes are located mostly in the inner part, where the epicyclic frequency $\kappa$ is maximal. Due to these spatial variations, the two bands of frequency of these two families of modes superimpose significantly, but separate asymptotically at $k_z \to 0$ and $k_z \to \infty$. Indeed, at low $k_z$, acoustic modes reach non-zero frequencies, while inertial waves all have zero frequency. On the other hand at large $k_z$, the inertial wave reach finite frequencies, while the acoustic waves follow $\omega \propto k_z$. \\
One mode falls outside this classification: it reaches zero frequency as $k_z \to 0$, yet behaves as $\omega \propto k_z$ for large $k_z$. This mode corresponds to the spectral flow: as $k_z$ increases, its branch transits from the inertial waveband to the acoustic waveband. Numerical integration shows that this branch corresponds to the fundamental modes of the disc, which have  no radial node of pressure or vertical velocity. It follows the dispersion relation $\omega = c_\mathrm{s}(r_0)k_z$ (grey thin line on Fig.~\ref{fig:monotonic}), as it is indeed located at the inner boundary $r_0$, and is acoustic in nature.

\begin{figure}
    \centering
    \includegraphics[width=0.7\columnwidth]{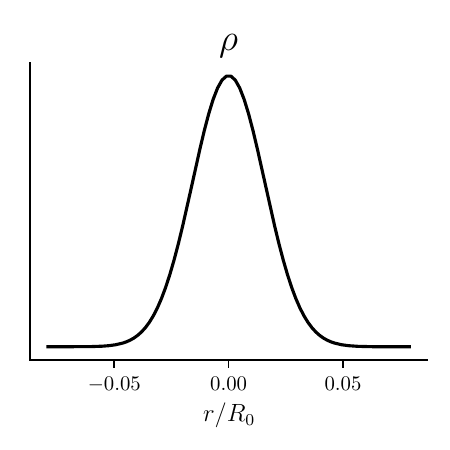}    
    \includegraphics[width=0.9\columnwidth]{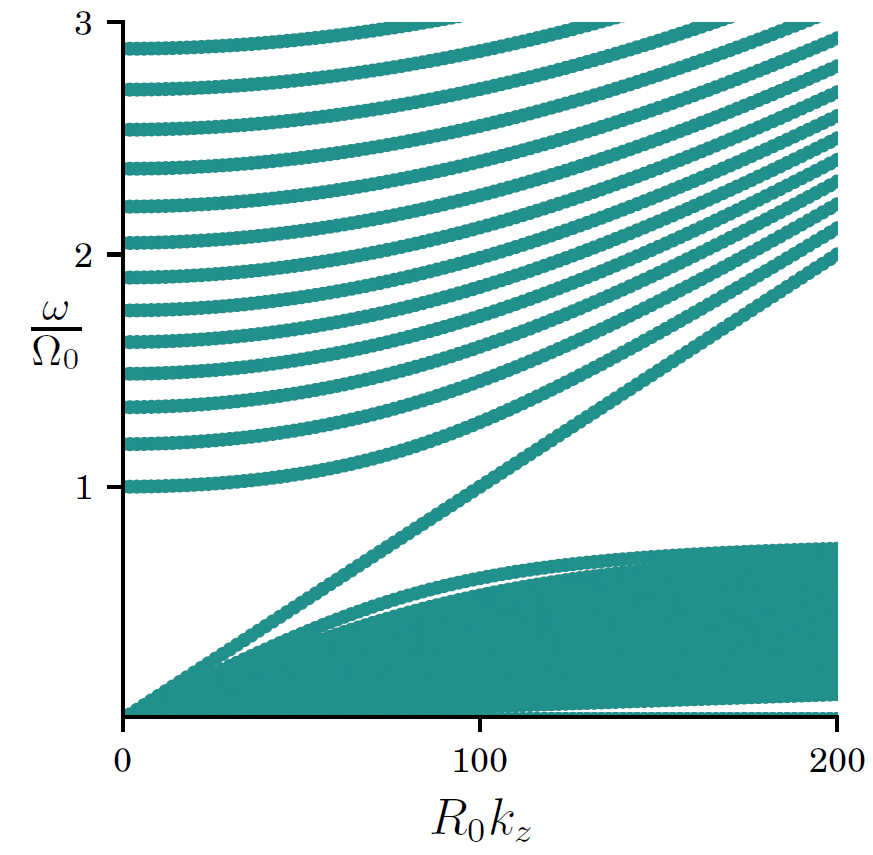}

    \caption{Dispersion relations of the global modes in a slender torus.}
    \label{fig:slenderTorus}
\end{figure}

\begin{figure}
    \centering
    \includegraphics[width=0.7\columnwidth]{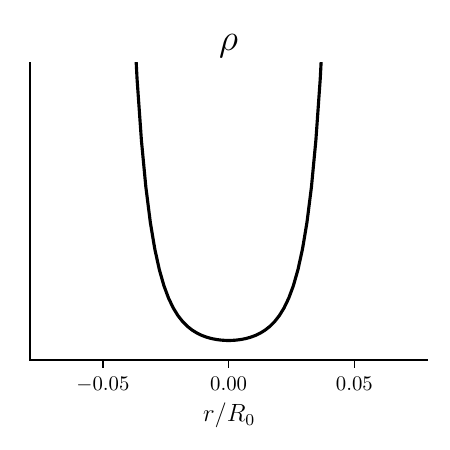}
    \includegraphics[width=0.9\columnwidth]{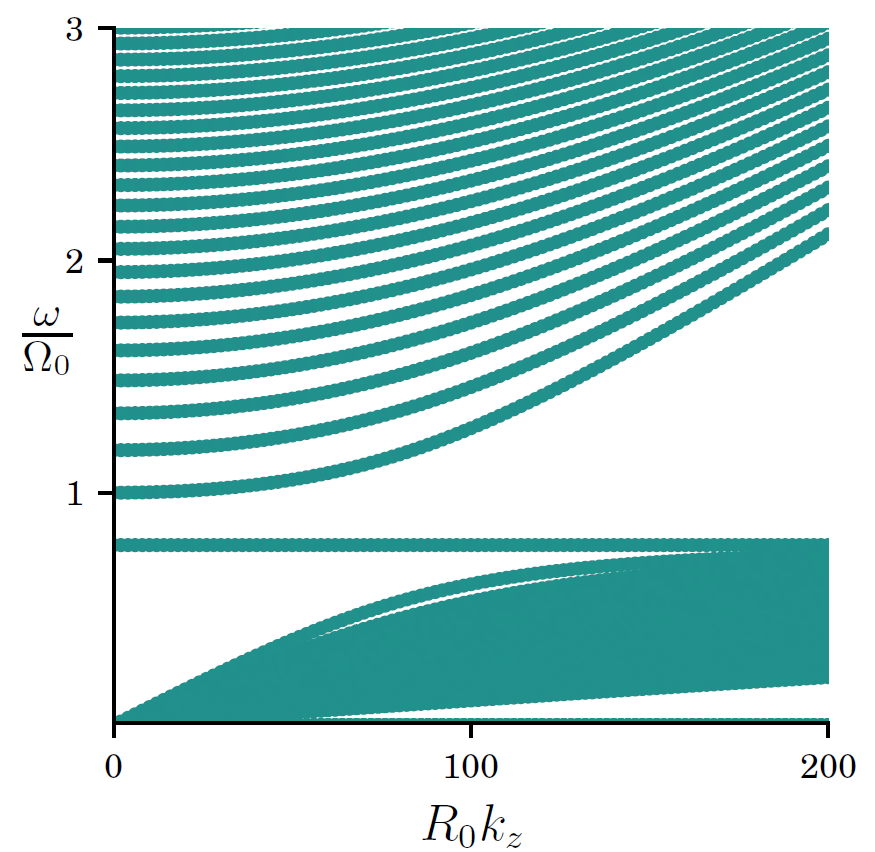}

    \caption{Dispersion relations of the global modes in a pressure minimum.}
    \label{fig:pressure-min}
\end{figure}

\vfill
\section{Pressure extrema}
\label{sec:extrema}

\subsection{Slender tori}
\label{subsec:slender}

Singularities of Berry curvature occur when $S=0$. Interestingly, when the spatial curvature term is negligible in Eq.~\eqref{eq:Salt}, this coincides with local pressure extrema, which occur in structured discs in the form of rings and gaps \citep{andrews2018}. One therefore expects topological modes to propagate around pressure maxima and minima. To investigate this expectation, we study the isothermal slender torus model \citep{papaloizou1985}, a model of rings in which the radial extent of the gas is assumed to be small compared to the distance to the star $r$, thereby enforcing strong radial pressure gradients. The gas density of the slender torus is parametrized by
\begin{eqnarray}
    \rho_0(r) &=& \rho_\mathrm{c} \exp\left(-\frac{(r-r_0)^2}{2L^2}\right), \label{eq:ST} \\
    S(r) &=& -\frac{c_\mathrm{s}}{2L^2}(r-r_0),\label{eq:S_STmax}
\end{eqnarray}
where here $r_0$ is the position of the pressure maximum, $L^2 = \frac{c_\mathrm{s}^2 \Omega_0^{-2}}{(2s-3)}$, with ${s = \frac{\dd \ln \Omega}{\dd \ln r}}$ and ${\Omega_0 = \Omega(r_0)}$. $S$ vanishes at $r = r_0$, which implies that the topological mode is trapped around this position.

Since $S$ varies linearly with $r$, analytical solutions for global modes can be found (a similar solution has been derived for stars in \citealt{leclerc2022}). For an isothermal disc, eliminating $u_\theta$ and $u_z$ in Eq.~\eqref{eq:eigenvalGlobalwaves}, it can then be written in the compact form 
\begin{eqnarray}
    u_r &=& \left(\omega-\frac{\kappa^2}{\omega}\right)^{-1}\mathcal{D}h,\\
    \mathcal{D}^\dagger\mathcal{D}h &=& \lambda h\label{eq:eigenvalueODEST_max},
\end{eqnarray}
where $\mathcal{D} \equiv ic_\mathrm{s}\partial_r -iS(r)$ and $\lambda \equiv \left(\omega-\frac{\kappa^2}{\omega}\right)\left(\omega-\frac{c_\mathrm{s}^2 k_z^2}{\omega}\right)$. The slender torus model verifies Eq.~\eqref{eq:S_STmax} and as such, the global modes equation Eq.~\eqref{eq:eigenvalueODEST_max} reads 
\begin{equation}
    \left(-\partial_{rr} + \frac{(r-r_0)^2}{4L^4} - \frac{1}{2L^2} - \frac{\lambda}{c_\mathrm{s}^2}\right)h = 0,\label{eq:ODE_QHO}
\end{equation}
which is the differential eigenvalue equation of the Quantum Harmonic Oscillator (\citealt{abramowitz1972}). Imposing regularity at infinity, the solutions of Eq.~\eqref{eq:ODE_QHO} are Hermite polynomials $H_n$ with associated frequencies $\omega_n$, so that the normal modes of the disc are 
\begin{eqnarray}
    p^\prime/\rho_0 &\propto& v_z^\prime \propto  H_n\left(\frac{r-r_0}{\sqrt{2}L}\right),\\
    v_r^\prime &\propto& v_\theta^\prime \propto \sqrt{n} H_{n-1}\left(\frac{r-r_0}{\sqrt{2}L}\right),\\
    \frac{n}{L^2} &=& \frac{(c_\mathrm{s}^2k_z^2-\omega_n^2)(\kappa_\mathrm{c}^2-\omega_n^2)}{c_\mathrm{s}^2 \omega_n^2}.
    \label{eq:freqsST_max}
\end{eqnarray}
In the solutions, $n$ is a non-negative integer that corresponds to the number of radial nodes in the pressure perturbation profile. For $n \geq 1$, two modes satisfy Eq.~\eqref{eq:freqsST_max}: one acoustic and one inertial, both of radial order $n$. For $n=0$, only a single solution with finite kinetic energy exists (for which $v_r^\prime = v_\theta^\prime=0$), namely
\begin{equation}
    \omega_{n=0} = \pm c_\mathrm{s}k_z.
\end{equation}
This mode propagates at all frequencies and as such, transits between the frequency bands. This mode and its behavior among the rest of the spectrum of the slender torus are shown on Fig.~\ref{fig:slenderTorus}.

\subsection{Pressure minima}
\label{subsec:pressure-dip}

Interestingly, the topological arguments also suggests modes trapped at a pressure minima. We therefore study a model of a gap given by the inverse of the slender torus profile, parametrized by
\begin{eqnarray}
    \rho_0(r) &=& \rho_\mathrm{c} \exp\left(+\frac{(r-r_0)^2}{2L^2}\right), \\
    S(r) &=& +\frac{c_\mathrm{s}}{2L^2}(r-r_0).
\end{eqnarray}
While $S$ passes through zero with a negative slope at pressure maxima, at pressure minima this slope is instead positive. Nevertheless, the differential equation Eq.\eqref{eq:globalWaves} differs very little, and the solutions of this model read
\begin{eqnarray}
    p^\prime/\rho_0 &\propto& v_z^\prime \propto \sqrt{n} \exp\left(-\frac{(r-r_0)^2}{2L^2}\right) H_{n-1}\left(\frac{r-r_0}{\sqrt{2}L}\right),\\
    v_r^\prime &\propto& v_\theta^\prime \propto \exp\left(-\frac{(r-r_0)^2}{2L^2}\right) H_{n}\left(\frac{r-r_0}{\sqrt{2}L}\right),\\
    \frac{n}{L^2} &=& \frac{(c_\mathrm{s}^2k_z^2-\omega_n^2)(\kappa_\mathrm{c}^2-\omega_n^2)}{c_\mathrm{s}^2 \omega_n^2} .
    \label{eq:freqsST_min}
\end{eqnarray}
Again, in this set of solutions, the $n=0$ case only has one solution with finite kinetic energy, which is the mode with constant frequency
\begin{equation}
    \omega = \pm \kappa
\end{equation}
This mode is a spectral flow \textit{in the other direction} than at a pressure maximum, since the branch transits from the acoustic band to the inertial band as $k_z$ increases, and is shown in Fig.~\ref{fig:pressure-min}.

\subsection{Gap in a disc}
\label{subsec:monotonic+gap}

\begin{figure}
    \centering
    \includegraphics[width=\columnwidth]{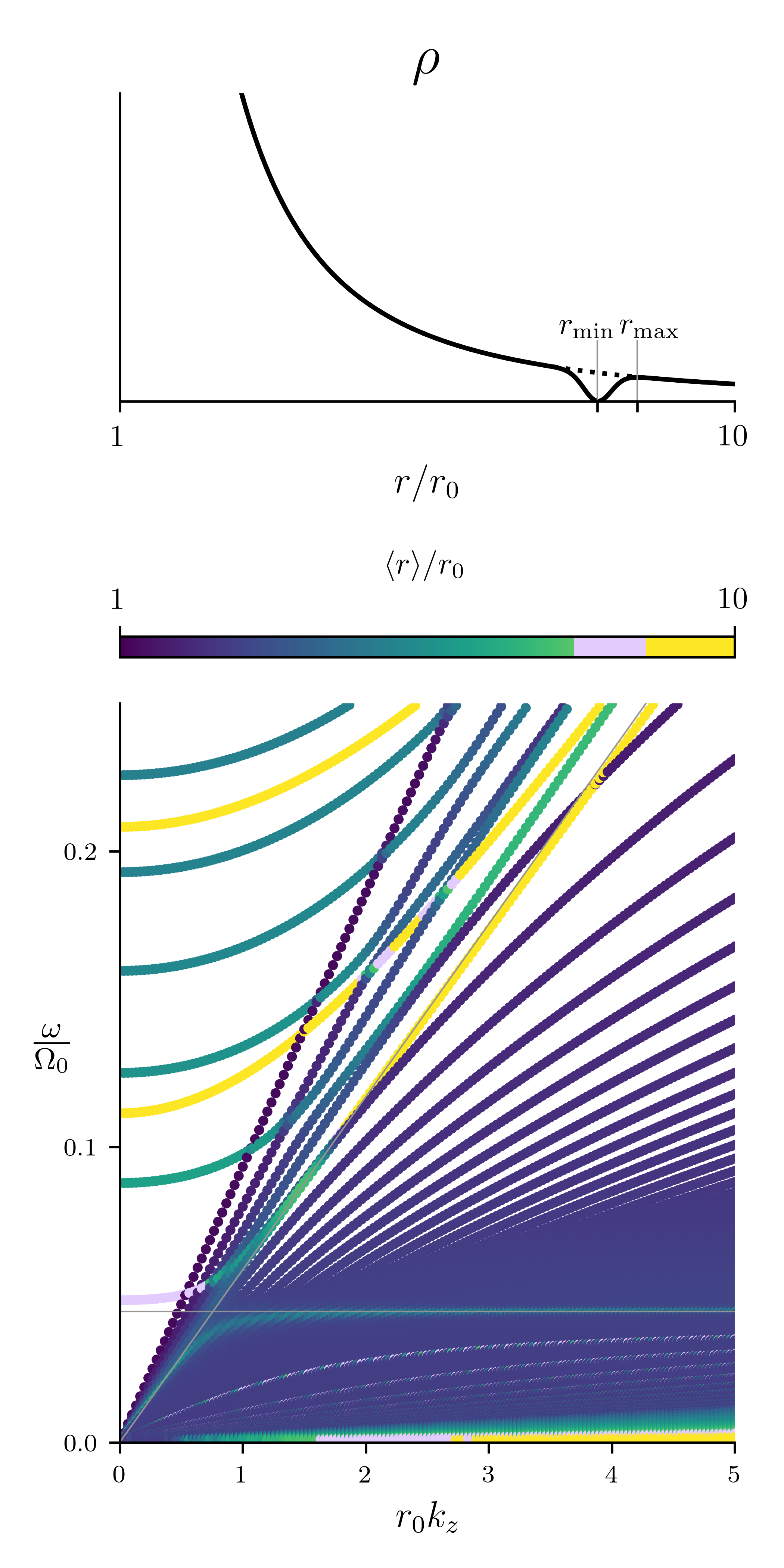}
    \caption{Same as Fig.~\ref{fig:monotonic}, for a disc with a gap and zoomed on lower frequencies. In a disc where a gap of density is present, one local minimum and one local maximum of density are found. Associated to these extrema, two topological modes are expected and found in the spectrum, one inertial and one acoustic respectively. Grey lines: $\omega = \kappa(r=r_\mathrm{min})$ and $\omega = c_\mathrm{s}(r=r_\mathrm{max})k_z$.}
    \label{fig:disc_with_gap}
\end{figure}

In order to test these predictions in a realistic gap in a disc, we study the spectrum of an extended disc with power-law profiles as discussed in Sec.~\ref{subsec:monotonic}, on top of which is carved a gap. More precisely, volume density is taken as
\begin{equation}
    \rho \propto r^{-p-s+\frac{q}{2}}\times\big(1-A\exp(-(r-r_\mathrm{gap})^2/2L_\mathrm{gap}^2)\big).
\end{equation}
This profile is shown on Fig.~\ref{fig:disc_with_gap} (top), with the position of the gap $r_\mathrm{gap} = 8 \,r_0$, width of the gap $L_\mathrm{gap}=0.2\, r_0$ and relative depth of the gap $A = 0.99$. \\
Interestingly, the addition of a gap in the disc causes the addition of two new extrema of density: a minimum and a maximum, separated by approximately half the gap width. By the analysis described in the sections above, one expects two topological modes localized at these regions of the disc, trapped by the density extrema. Close to $r_\mathrm{min}$ and therefore close to the center of the gap, one expects an epicyclic mode with dispersion relation $\omega = \kappa(r=r_\mathrm{min})$, and an acoustic mode close to $r_\mathrm{max}$ with dispersion relation $\omega = c_\mathrm{s}(r=r_\mathrm{max})k_z$.\\

The bottom panel of Fig. \ref{fig:disc_with_gap} presents the numerically obtained spectrum of the disc. As before, the colors indicate the average mode positions, with modes located inside the gap highlighted in pink.\\
In addition to the inertial and acoustic branches already present in the monotonic disc, one observes a new set of inertial branches (low-frequency, pink) as well as a new set of acoustic branches (yellow). Moreover, the two branches of modes predicted by the topological analysis are also visible, with their dispersion relations overlaid as thin grey lines.
An inertial branch appears at $\omega = \kappa(r = r_\mathrm{min})$, while an acoustic branch follows $\omega = c_\mathrm{s}(r = r_\mathrm{max})k_z$. Because these two modes occupy nearby regions in radius, they hybridize when their frequencies align, leading to an avoided crossing near $r_0 k_z \sim 0.8$. The low-frequency pink branch is not incidental: it represents a mode trapped in the gap and corresponds to the topological epicyclic mode, as evidenced by its eigenfunctions (see Appendix~\ref{app:profiles_modes}).\\
Therefore, the density extrema act as waveguides, trapping modes in the way described in the analytical models presented in Sect.~\ref{subsec:slender} and \ref{subsec:pressure-dip}. In addition to the two topological modes, new branches of inertial modes appeared in the spectrum at low frequency, trapped in the gap. New branches of acoustic modes appear with average position in the yellow region, as those are instead trapped between the gap and the outer boundary. In that case, it appears that the gap of the disc acts as a reflector for acoustic waves.

\section{Discussion}
\label{sec:discussion}
By leveraging topological methods, our analysis characterizes how large-scale modes behave in radially structured discs. Emphasis was put on the role that radial pressure and density gradients play on linear stable waves, encapsulated by the characteristic frequency $S$. Using microlocal techniques, we show that these gradients modify the local dispersion relations by coupling acoustic and inertial waves and removing their frequency degeneracy. As such, $S$ naturally acts as a cutoff frequency and offers a consistent way to account for gradient effects in local analyses. Topological arguments highlight the significance of points where $S=0$, which carry topological charges (Chern numbers $\mathcal{C}=\pm1$). These charges enforce the emergence of modes with unique spectral properties, namely branches that transit between $r$-modes and $p$-modes. While a monotonic disc hosts only one such branch, structured discs contain additional ones, as illustrated in Fig.~\ref{fig:disc_with_gap}. We emphasize that the gapped-disc model presented in Sect.~\ref{subsec:monotonic+gap} is intended as an illustrative example of trapped modes; although it reproduces the expected topological modes at pressure extrema, it is not meant to be realistic or to remain stable against the Rossby wave instability \citep{lovelace1999rossby}.\\

The fundamental mode trapped at a pressure maximum propagates for all frequencies and is the sole mode to do so (see Fig.~\ref{fig:slenderTorus}), making it capable of resonating with any temporal forcing, unlike modes with forbidden frequency domains. It is a vertical acoustic wave (see Figs.~\ref{fig:eigenprofiles_acousticMode_gap}-\ref{fig:eigenprofiles_acousticMode_gap_lowK}). By contrast, the fundamental mode associated with a pressure minimum (a gap) propagates only at the constant frequency $\omega = \kappa$ (see Fig.~\ref{fig:pressure-min} and Figs.~\ref{fig:eigenprofiles_gap_lowK}-\ref{fig:eigenprofiles_gap_highK}), and is a horizontal, epicyclic wave. This unique property thereby allows for a propagation at arbitrary vertical phase velocities. Indeed, since $\frac{\omega}{k_z} = \frac{\kappa}{k_z}$ can take any value, this mode is unique among the modes in having an arbitrary vertical phase velocity. This is a necessary condition for resonating with any vertical dust flow involved in dust-settling instabilities, a class of resonant drag instabilities (e.g. \citealt{squire2018,Zhuravlev2019,Lehmann2023,Paardekooper2025}). In the analytical model of Sec.~\ref{subsec:pressure-dip}, this mode has no vertical velocity and therefore does not couple to vertical flows \textit{a priori}. However, we emphasize that this is a peculiarity of that model. In general, the mode does exhibit some vertical-velocity perturbation (see Appendix~\ref{app:profiles_modes}): this topological mode may thus couple to vertical flows in settling-instability mechanisms. It would be worthwhile to further investigate its role at the outer edge of a planetary gap, where dust accumulates.\\
The spectrum of a disc with a pressure bump instead of a pressure gap produces similar results (see Fig.~\ref{fig:eigenprofiles_bump}).\\

At low $k_z$ (see Figure \ref{fig:eigenprofiles_gap_lowK}) this mode exhibits a very similar character to the trapped $r$-modes found in black hole accretion discs \citep{Kato2004,Kato2008,Ferreira2008,Dewberry2018}. Only here the mode is confined by the epicyclic-acoustic frequency rather than by the maximum in the relativistic epicyclic frequency found in black hole discs. This leaves open the possibility that this mode could be excited to large amplitudes via a similar three-wave coupling mechanism \citep{Kato2004,Kato2008,Ferreira2008}, where the $r$-mode is coupled to a global $m=1$ deformation in such a way that it leads to growth in the $r$-mode amplitude. Candidates for such a global $m=1$ deformation include remnant disc eccentricity left over from the disc formation process \citep{Commercon2024}, or disc warping and eccentricity excited by an embedded planet \citep{Lubow2001,Teyssandier2016}.\\

The main limitations of the present study are the absence of vertical stratification and of any linear instability. The vertical stratification is sometimes separated from the radial problem by averaging vertically (e.g. \citealt{papaloizou1985} or \citealt{blaes2006}); we instead chose to assume no stratification, in order to keep the notion of a vertical wavenumber $k_z$. We note that our analytical solutions for the slender torus Eq.~\eqref{eq:freqsST_max} matches those of \citet{blaes2006}, see their Eq.~56 for their $n$ set to $\infty$ and our $k_z=0$. \\
We also note that Eq.~\eqref{eq:localDisp} is analogous to the one found by \citealt{okazaki1987} (Eq. 4.4) who studied vertical isothermal stratification and no radial stratification, with two expected differences: the addition of $S^2$ representing radial gradients in our case, and the replacement of $c_\mathrm{s}^2k_z^2$ by $n_z \kappa^2$ in their case, where $n_z$ is a positive integer counting the number of nodes in the profile of pressure perturbation given by the Hermite function of order $n_z$. Thus, vertical isothermal stratification seems to be fully taken into account by discretizing wavenumbers as $c_\mathrm{s}k_z = n_z^{1/2}\kappa$. However, the results of \citealt{korycansky1995} suggest an interesting behavior of the modes with vertical stratification and no radial gradients (see their Fig.1). Indeed, they show that a pair of fundamental $p$-modes with indices $n_z = 0,1$ have frequencies connecting inertial modes and acoustic modes, possibly performing a spectral flow. This suggests that, complementarily, vertical stratification imposes another set of topological properties and topological modes, regardless of radial gradients.\\
The analytical techniques and solutions found for modes of non-axisymmetric polytropic discs, slender \citep{papaloizou1985,blaes1985,blaes2006} or thick \citep{blaes2007} provide the support to explore the topology of these waves rising from vertical inhomogeneity, and from azimuthal wavenumbers, and possibly associated topological modes. Our goal in the present study is to provide a starting point for applying topological analysis to linear waves in discs, a framework whose predictive power can yield new insights into large-scale oscillations in inhomogeneous discs, with the perspective of establishing robust and possibly unique properties of large-scale modes falling out of WKB approaches. This line of research is motivated by the anticipated potential of discoseismology in protoplanetary discs, facilitated by line-kinematic measurements from ALMA (e.g. \citealt{Pinte2023,Teague2025}). While this perspective is still far off, we expect that the spatial patterns of the eigenmodes may provide constraints on models—a key distinction from stellar pulsation studies. The employment of topological analysis to study linear instabilities in discs requires further investigation, as it is a field of research less developed than topology of stable waves. In stellar context, these ideas have led to the unveiling of fast-growing large-scale modes of compressible convective instability \citep{Leclerc2024a}. Similar results may be expected in discs instabilities, when accounting for the role of pressure gradients as we did in the present study.

\section{Conclusion}
Motivated by the foreseen potential of discoseismology in protoplanetary discs enabled by line–kinematic measurements from ALMA, we conduct a topological analysis of radial waves in unstratified astrophysical discs. The main conclusions are:
\begin{enumerate}
\item \textbf{Existence of robust topological modes} : Global large-scale modes transiting between the inertial and the acoustic bands exist in discs. These modes find their roots in topological numbers, and as such qualify as being \textit{topological modes}. This special origin provides them with unique properties.
\item  \textbf{Epicyclic–acoustic frequency} : The existence of topological modes in discs is associated with the cancellation of the \textit{epicyclic-acoustic} frequency $S$ given by Eq.~\ref{eq:S}, which gives the rate of momentum exchange between the inertial and acoustic bands. This is why monotonic-density discs support a single topological mode at their inner edge.
\item \textbf{Role of pressure extrema} : Pressure maxima and minima, acting as waveguides, generate additional topological modes because $S = 0$ at these locations. The associated topological branch is acoustic at pressure maxima and inertial at pressure minima, both having unique specificity across the spectrum (any frequency or any phase velocity). Their distinct features make them appealing targets for future line-based discoseismology. Furthermore, they are potentially easier to detect as they are localized, and their radial trapping would provide a direct measure of the pressure gradient of the bump/dip.
\item \textbf{Microlocal framework} : The microlocal analysis introduced here enables the study of local wave properties without using a shearing box, while preserving the Hermitian structure of the problem. This approach provides a powerful framework for studying linear modes in discs.
\end{enumerate}
As commented on in the discussion, we focused on the midplane, ignoring for the moment the influence of vertical stratification. Treating simultaneously vertical and radial gradients requires a complementary study, which may reveal additional topological structures in the higher-dimensional parameter space involved. A stepping stone towards general results and potentially discoseismology is to study a dual simplified problem where radial gradients are neglected while keeping vertical stratification. A study focusing on the topological modes of this problem will be the object of a future work.  

\section*{Acknowledgements}
AL was funded by Contrat Doctoral Spécifique Normaliens during this work. GL, EL and NP acknowledge funding from ERC CoG project
PODCAST No 864965. We used \textsc{Mathematica} \citep{Mathematica}. The scripts used for numerical calculations are accessible at \url{https://github.com/ArmandLeclerc/radialStratifDisc}.

\bibliographystyle{unsrtnat}
\bibliography{radialStratif}{}

@article{latter2017local,
  title={Local models of astrophysical discs},
  author={Latter, Henrik N and Papaloizou, John},
  journal={Monthly Notices of the Royal Astronomical Society},
  volume={472},
  number={2},
  pages={1432--1446},
  year={2017},
  publisher={Oxford University Press}
}

@article{goldreich1965,
  title={II. Spiral arms as sheared gravitational instabilities},
  author={Goldreich, Peter and Lynden-Bell, D},
  journal={Monthly Notices of the Royal Astronomical Society},
  volume={130},
  number={2},
  pages={125--158},
  year={1965},
  publisher={Oxford University Press Oxford, UK}
}

@article{hill1878,
  title={Researches in the lunar theory},
  author={Hill, George William},
  journal={American journal of Mathematics},
  volume={1},
  number={1},
  pages={5--26},
  year={1878},
  publisher={JSTOR}
}

@article{balbus1991,
  title={A powerful local shear instability in weakly magnetized disks. I-Linear analysis. II-Nonlinear evolution},
  author={Balbus, Steven A and Hawley, John F},
  journal={Astrophysical Journal, Part 1 (ISSN 0004-637X), vol. 376, July 20, 1991, p. 214-233.},
  volume={376},
  pages={214--233},
  year={1991}
}

@ARTICLE{Commercon2024,
       author = {{Commer{\c{c}}on}, Beno{\^\i}t and {Lovascio}, Francesco and {Lynch}, Elliot and {Ragusa}, Enrico},
        title = "{Discs are born eccentric}",
      journal = {\aap},
         year = 2024,
        month = sep,
       volume = {689},
          eid = {L9},
        pages = {L9},
          doi = {10.1051/0004-6361/202449610},
archivePrefix = {arXiv},
       eprint = {2408.16319},
 primaryClass = {astro-ph.SR},
       adsurl = {https://ui.adsabs.harvard.edu/abs/2024A&A...689L...9C}}

@article{youdin2005,
  title={Streaming instabilities in protoplanetary disks},
  author={Youdin, Andrew N and Goodman, Jeremy},
  journal={The Astrophysical Journal},
  volume={620},
  number={1},
  pages={459},
  year={2005},
  publisher={IOP Publishing}
}

@article{nelson2013,
  title={Linear and non-linear evolution of the vertical shear instability in accretion discs},
  author={Nelson, Richard P and Gressel, Oliver and Umurhan, Orkan M},
  journal={Monthly Notices of the Royal Astronomical Society},
  volume={435},
  number={3},
  pages={2610--2632},
  year={2013},
  publisher={Oxford University Press}
}

@article{johansen2007,
  title={Rapid planetesimal formation in turbulent circumstellar disks},
  author={Johansen, Anders and Oishi, Jeffrey S and Low, Mordecai-Mark Mac and Klahr, Hubert and Henning, Thomas and Youdin, Andrew},
  journal={Nature},
  volume={448},
  number={7157},
  pages={1022--1025},
  year={2007},
  publisher={Nature Publishing Group UK London}
}

@article{balbus1998,
  title={Instability, turbulence, and enhanced transport in accretion disks},
  author={Balbus, Steven A and Hawley, John F},
  journal={Reviews of modern physics},
  volume={70},
  number={1},
  pages={1},
  year={1998},
  publisher={APS}
}

@book{hall2013,
  title={Quantum theory for mathematicians},
  author={Hall, Brian C},
  year={2013},
  publisher={Springer}
}

@book{volovik2003,
  title={The universe in a helium droplet},
  author={Volovik, Grigory E},
  volume={117},
  year={2003},
  publisher={OUP Oxford}
}

@article{delplace2022,
  title={Berry-Chern monopoles and spectral flows},
  author={Delplace, Pierre},
  journal={SciPost Physics Lecture Notes},
  pages={039},
  year={2022}
}

@article{armitage2011,
  title={Dynamics of protoplanetary disks},
  author={Armitage, Philip J},
  journal={Annual Review of Astronomy and Astrophysics},
  volume={49},
  pages={195--236},
  year={2011},
  publisher={Annual Reviews}
}

@article{lin2015,
  title={Cooling requirements for the vertical shear instability in protoplanetary disks},
  author={Lin, Min-Kai and Youdin, Andrew N},
  journal={The Astrophysical Journal},
  volume={811},
  number={1},
  pages={17},
  year={2015},
  publisher={IOP Publishing}
}

@article{onuki2020,
  title={Quasi-local method of wave decomposition in a slowly varying medium},
  author={Onuki, Yohei},
  journal={Journal of Fluid Mechanics},
  volume={883},
  pages={A56},
  year={2020},
  publisher={Cambridge University Press}
}

@article{leclerc2022,
  title={Topological modes in stellar oscillations},
  author={Leclerc, Armand and Laibe, Guillaume and Delplace, Pierre and Venaille, Antoine and Perez, Nicolas},
  journal={The Astrophysical Journal},
  volume={940},
  number={1},
  pages={84},
  year={2022},
  publisher={IOP Publishing}
}

@article{perrot2019,
  title={Topological transition in stratified fluids},
  author={Perrot, Manolis and Delplace, Pierre and Venaille, Antoine},
  journal={Nature Physics},
  volume={15},
  number={8},
  pages={781--784},
  year={2019},
  publisher={Nature Publishing Group UK London}
}

@article{delplace2017,
  title={Topological origin of equatorial waves},
  author={Delplace, Pierre and Marston, JB and Venaille, Antoine},
  journal={Science},
  volume={358},
  number={6366},
  pages={1075--1077},
  year={2017},
  publisher={American Association for the Advancement of Science}
}

@article{perez2022,
  title={Unidirectional modes induced by nontraditional Coriolis force in stratified fluids},
  author={Perez, Nicolas and Delplace, Pierre and Venaille, Antoine},
  journal={Physical Review Letters},
  volume={128},
  number={18},
  pages={184501},
  year={2022},
  publisher={APS}
}

@article{burns2020,
  title={Dedalus: A flexible framework for numerical simulations with spectral methods},
  author={Burns, Keaton J and Vasil, Geoffrey M and Oishi, Jeffrey S and Lecoanet, Daniel and Brown, Benjamin P},
  journal={Physical Review Research},
  volume={2},
  number={2},
  pages={023068},
  year={2020},
  publisher={APS}
}

@misc{Mathematica,
  author = {{Wolfram Research{,} Inc.}},
  title = {Mathematica, {V}ersion 14.0},
  url = {https://www.wolfram.com/mathematica},
  note = {Champaign, IL},
  year = {2024}
}

@article{qin2023,
  title={Topological Langmuir-cyclotron wave},
  author={Qin, Hong and Fu, Yichen},
  journal={Science Advances},
  volume={9},
  number={13},
  pages={eadd8041},
  year={2023},
  publisher={American Association for the Advancement of Science}
}

@article{parker2020,
  title={Topological gaseous plasmon polariton in realistic plasma},
  author={Parker, Jeffrey B and Marston, JB and Tobias, Steven M and Zhu, Ziyan},
  journal={Physical Review Letters},
  volume={124},
  number={19},
  pages={195001},
  year={2020},
  publisher={APS}
}

@article{papaloizou1985,
  title={The dynamical stability of differentially rotating discs--II},
  author={Papaloizou, JCB and Pringle, JE},
  journal={Monthly Notices of the Royal Astronomical Society},
  volume={213},
  number={4},
  pages={799--820},
  year={1985},
  publisher={Oxford University Press Oxford, UK}
}

@article{oishi2021,
  title={eigentools: A Python package for studying differential eigenvalue problems with an emphasis on robustness},
  author={Oishi, Jeffrey S and Burns, Keaton J and Clark, Susan E and Anders, Evan H and Brown, Benjamin P and Vasil, Geoffrey M and Lecoanet, Daniel},
  journal={Journal of Open Source Software},
  volume={6},
  number={62},
  pages={3079},
  year={2021}
}

@misc{keppeler2004,
    title={Introduction to Wigner-Weyl calculus},
    author={Keppeler,Stefan},
    url ={https://www.math.uni-tuebingen.de/user/stke/teaching/wigner\_weyl},
    year ={2004}
}

@article{goldreich1965a,
  title={I. Gravitational stability of uniformly rotating disks},
  author={Goldreich, Peter and Lynden-Bell, D},
  journal={Monthly Notices of the Royal Astronomical Society},
  volume={130},
  number={2},
  pages={97--124},
  year={1965},
  publisher={Oxford University Press Oxford, UK}
}

@article{lynden1972,
  title={On the generating mechanism of spiral structure},
  author={Lynden-Bell, D and Kalnajs, AJ},
  journal={Monthly Notices of the Royal Astronomical Society},
  volume={157},
  number={1},
  pages={1--30},
  year={1972},
  publisher={Oxford University Press Oxford, UK}
}

@article{goldreich1978,
  title={The excitation and evolution of density waves},
  author={Goldreich, Peter and Tremaine, Scott},
  journal={Astrophysical Journal},
  volume={222},
  number={1},
  pages={850--858},
  year={1978},
  publisher={American Astronomical Society}
}

@article{goldreich1979,
  title={The excitation of density waves at the Lindblad and corotation resonances by an external potential},
  author={Goldreich, Peter and Tremaine, Scott},
  journal={Astrophysical Journal},
  volume={233},
  number={3},
  pages={857--871},
  year={1979},
  publisher={American Astronomical Society}
}

@article{korycansky1995,
  title = {Axisymmetric Waves in Polytropic Accretion Discs},
  author = {Korycansky, D. G. and Pringle, J. E.},
  year = {1995},
  month = feb,
  journal = {Monthly Notices of the Royal Astronomical Society},
  volume = {272},
  number = {3},
  pages = {618--624},
  issn = {0035-8711},
  doi = {10.1093/mnras/272.3.618},
  urldate = {2024-01-27}
}

@article{lubow1993,
  title = {Wave {{Propagation}} in {{Accretion Disks}}: {{Axisymmetric Case}}},
  shorttitle = {Wave {{Propagation}} in {{Accretion Disks}}},
  author = {Lubow, S. H. and Pringle, J. E.},
  year = {1993},
  month = may,
  journal = {The Astrophysical Journal},
  volume = {409},
  pages = {360},
  issn = {0004-637X},
  doi = {10.1086/172669},
  urldate = {2024-01-27}
}

@article{lubow1998,
  title = {Three-Dimensional {{Waves Generated}} at {{Lindblad Resonances}} in {{Thermally Stratified Disks}}},
  author = {Lubow, S. H. and Ogilvie, G. I.},
  year = {1998},
  month = sep,
  journal = {The Astrophysical Journal},
  volume = {504},
  number = {2},
  pages = {983},
  issn = {0004-637X},
  doi = {10.1086/306104},
  urldate = {2024-01-27}
}

@ARTICLE{Lubow2001,
       author = {{Lubow}, S.~H. and {Ogilvie}, G.~I.},
        title = "{Secular Interactions between Inclined Planets and a Gaseous Disk}",
      journal = {\apj},
         year = 2001,
        month = oct,
       volume = {560},
       number = {2},
        pages = {997-1009},
          doi = {10.1086/322493},
archivePrefix = {arXiv},
       eprint = {astro-ph/0106453},
 primaryClass = {astro-ph},
       adsurl = {https://ui.adsabs.harvard.edu/abs/2001ApJ...560..997L}}

@article{ogilvie1998,
  title = {Waves and Instabilities in a Differentially Rotating Disc Containing a Poloidal Magnetic Field},
  author = {Ogilvie, G.I.},
  year = {1998},
  month = jun,
  journal = {Monthly Notices of the Royal Astronomical Society},
  volume = {297},
  number = {1},
  pages = {291--314},
  issn = {0035-8711},
  doi = {10.1046/j.1365-8711.1998.01507.x},
  urldate = {2024-01-27}
}

@article{ogilvie1999,
  title = {The {{Effect}} of an {{Isothermal Atmosphere}} on the {{Propagation}} of {{Three-dimensional Waves}} in a {{Thermally Stratified Accretion Disk}}},
  author = {Ogilvie, G. I. and Lubow, S. H.},
  year = {1999},
  month = apr,
  journal = {The Astrophysical Journal},
  volume = {515},
  number = {2},
  pages = {767},
  issn = {0004-637X},
  doi = {10.1086/307037},
  urldate = {2024-01-27}
}

@ARTICLE{Ferreira2008,
       author = {{Ferreira}, B{\'a}rbara T. and {Ogilvie}, Gordon I.},
        title = "{On an excitation mechanism for trapped inertial waves in discs around black holes}",
      journal = {\mnras},
         year = 2008,
        month = jun,
       volume = {386},
       number = {4},
        pages = {2297-2310},
          doi = {10.1111/j.1365-2966.2008.13207.x},
archivePrefix = {arXiv},
       eprint = {0803.1671},
 primaryClass = {astro-ph},
       adsurl = {https://ui.adsabs.harvard.edu/abs/2008MNRAS.386.2297F}}

@article{hedman2013,
  title={Kronoseismology: using density waves in Saturn's C ring to probe the planet's interior},
  author={Hedman, MM and Nicholson, PD},
  journal={The Astronomical Journal},
  volume={146},
  number={1},
  pages={12},
  year={2013},
  publisher={IOP Publishing}
}

@ARTICLE{Teyssandier2016,
       author = {{Teyssandier}, Jean and {Ogilvie}, Gordon I.},
        title = "{Growth of eccentric modes in disc-planet interactions}",
      journal = {\mnras},
         year = 2016,
        month = may,
       volume = {458},
       number = {3},
        pages = {3221-3247},
          doi = {10.1093/mnras/stw521},
archivePrefix = {arXiv},
       eprint = {1603.00653},
 primaryClass = {astro-ph.EP},
       adsurl = {https://ui.adsabs.harvard.edu/abs/2016MNRAS.458.3221T}}

@article{toomre1964,
  title={On the gravitational stability of a disk of stars},
  author={Toomre, Alar},
  journal={Astrophysical Journal, vol. 139, p. 1217-1238 (1964).},
  volume={139},
  pages={1217--1238},
  year={1964}
}

@misc{vidal2024,
      title={Inertia-gravity waves in geophysical vortices}, 
      author={Jérémie Vidal and Yves Colin de Verdière},
      year={2024},
      eprint={2402.10749},
      archivePrefix={arXiv},
      primaryClass={physics.flu-dyn}
}

@article{faure2023,
  title={Manifestation of the topological index formula in quantum waves and geophysical waves},
  author={Faure, Fr{\'e}d{\'e}ric},
  journal={Annales Henri Lebesgue},
  volume={6},
  pages={449--492},
  year={2023}
}

@ARTICLE{squire2018,
       author = {{Squire}, Jonathan and {Hopkins}, Philip F.},
        title = "{Resonant drag instabilities in protoplanetary discs: the streaming instability and new, faster growing instabilities}",
      journal = {\mnras},
         year = 2018,
        month = jul,
       volume = {477},
       number = {4},
        pages = {5011-5040},
          doi = {10.1093/mnras/sty854},
archivePrefix = {arXiv},
       eprint = {1711.03975},
 primaryClass = {astro-ph.EP},
       adsurl = {https://ui.adsabs.harvard.edu/abs/2018MNRAS.477.5011S}
}

@ARTICLE{lyndenBell1967,
       author = {{Lynden-Bell}, D. and {Ostriker}, J.~P.},
        title = "{On the stability of differentially rotating bodies}",
      journal = {\mnras},
         year = 1967,
        month = jan,
       volume = {136},
        pages = {293},
          doi = {10.1093/mnras/136.3.293},
       adsurl = {https://ui.adsabs.harvard.edu/abs/1967MNRAS.136..293L}
}

@ARTICLE{lindblad1948,
       author = {{Lindblad}, Bertil},
        title = "{On the Dynamics of Stellar Systems (George Darwin Lecture)}",
      journal = {\mnras},
         year = 1948,
        month = jan,
       volume = {108},
        pages = {214},
          doi = {10.1093/mnras/108.3.214},
       adsurl = {https://ui.adsabs.harvard.edu/abs/1948MNRAS.108..214L}
}

@ARTICLE{lin1966,
       author = {{Lin}, C.~C. and {Shu}, Frank H.},
        title = "{On the Spiral Structure of Disk Galaxies, II. Outline of a Theory of Density Waves}",
      journal = {Proceedings of the National Academy of Science},
         year = 1966,
        month = feb,
       volume = {55},
       number = {2},
        pages = {229-234},
          doi = {10.1073/pnas.55.2.229},
       adsurl = {https://ui.adsabs.harvard.edu/abs/1966PNAS...55..229L}
}

@ARTICLE{goldreich1980,
       author = {{Goldreich}, P. and {Tremaine}, S.},
        title = "{Disk-satellite interactions.}",
      journal = {\apj},
         year = 1980,
        month = oct,
       volume = {241},
        pages = {425-441},
          doi = {10.1086/158356},
       adsurl = {https://ui.adsabs.harvard.edu/abs/1980ApJ...241..425G}
}

@ARTICLE{goodman1993,
       author = {{Goodman}, Jeremy},
        title = "{A Local Instability of Tidally Distorted Accretion Disks}",
      journal = {\apj},
         year = 1993,
        month = apr,
       volume = {406},
        pages = {596},
          doi = {10.1086/172472},
       adsurl = {https://ui.adsabs.harvard.edu/abs/1993ApJ...406..596G}
}

@ARTICLE{kato1978,
       author = {{Kato}, Shoji},
        title = "{Pulsational instability of accretion disks to axially symmetric oscillations}",
      journal = {\mnras},
         year = 1978,
        month = dec,
       volume = {185},
        pages = {629-642},
          doi = {10.1093/mnras/185.3.629},
       adsurl = {https://ui.adsabs.harvard.edu/abs/1978MNRAS.185..629K}
}

@ARTICLE{Kato2004,
       author = {{Kato}, Shoji},
        title = "{Resonant Excitation of Disk Oscillations by Warps: A Model of kHz QPOs}",
      journal = {\pasj},
         year = 2004,
        month = oct,
       volume = {56},
        pages = {905-922},
          doi = {10.1093/pasj/56.5.905},
archivePrefix = {arXiv},
       eprint = {astro-ph/0409051},
 primaryClass = {astro-ph},
       adsurl = {https://ui.adsabs.harvard.edu/abs/2004PASJ...56..905K}}

@ARTICLE{Kato2008,
       author = {{Kato}, Shoji},
        title = "{Resonant Excitation of Disk Oscillations in Deformed Disks II: A Model of High-Frequency QPOs}",
      journal = {\pasj},
         year = 2008,
        month = feb,
       volume = {60},
        pages = {111},
          doi = {10.1093/pasj/60.1.111},
archivePrefix = {arXiv},
       eprint = {0709.2467},
 primaryClass = {astro-ph},
       adsurl = {https://ui.adsabs.harvard.edu/abs/2008PASJ...60..111K}}

@article{blaes1985,
  title={Oscillations of slender tori},
  author={Blaes, OM},
  journal={Monthly Notices of the Royal Astronomical Society},
  volume={216},
  number={3},
  pages={553--563},
  year={1985},
  publisher={The Royal Astronomical Society}
}

@article{blaes2006,
  title={Oscillation modes of relativistic slender tori},
  author={Blaes, Omer M and Arras, P and Fragile, PC},
  journal={Monthly Notices of the Royal Astronomical Society},
  volume={369},
  number={3},
  pages={1235--1252},
  year={2006},
  publisher={Blackwell Publishing Ltd Oxford, UK}
}

@article{blaes2007,
  title={Epicyclic oscillations of fluid bodies: Newtonian nonslender torus},
  author={Blaes, Omer M and {\v{S}}r{\'a}mkov{\'a}, Eva and Abramowicz, Marek A and Klu{\'z}niak, W{\l}odek and Torkelsson, Ulf},
  journal={The Astrophysical Journal},
  volume={665},
  number={1},
  pages={642},
  year={2007},
  publisher={IOP Publishing}
}

@article{kato1983,
  title={Low-frequency, one-armed oscillations of Keplerian gaseous disks},
  author={Kato, Shoji},
  journal={Astronomical Society of Japan, Publications (ISSN 0004-6264), vol. 35, no. 2, 1983, p. 249-261.},
  volume={35},
  pages={249--261},
  year={1983}
}

@article{kato1980,
  title={Trapped radial oscillations of gaseous disks around a black hole},
  author={Kato, Shoji and Fukue, Jun},
  journal={Publications of the Astronomical Society of Japan, Vol. 32, P. 377, 1980},
  volume={32},
  pages={377},
  year={1980}
}

@BOOK{abramowitz1972,
       author = {{Abramowitz}, M. and {Stegun}, I.~A.},
        title = "{Handbook of Mathematical Functions}",
         year = 1972,
       adsurl = {https://ui.adsabs.harvard.edu/abs/1972hmfw.book.....A},
    publisher={US Department of Commerce}
}

@article{andrews2018,
  title={The disk substructures at high angular resolution project (DSHARP). I. Motivation, sample, calibration, and overview},
  author={Andrews, Sean M and Huang, Jane and P{\'e}rez, Laura M and Isella, Andrea and Dullemond, Cornelis P and Kurtovic, Nicol{\'a}s T and Guzm{\'a}n, Viviana V and Carpenter, John M and Wilner, David J and Zhang, Shangjia and others},
  journal={The Astrophysical Journal Letters},
  volume={869},
  number={2},
  pages={L41},
  year={2018},
  publisher={IOP Publishing}
}

@ARTICLE{LL2025,
       author = {{Le Saux}, Arthur and {Leclerc}, Armand and {Laibe}, Guillaume and {Delplace}, Pierre and {Venaille}, Antoine},
        title = "{A Core-sensitive Mixed f/g-mode of the Sun Predicted by Wave Topology and Hydrodynamical Simulation}",
      journal = {\apjl},
         year = 2025,
        month = jul,
       volume = {987},
       number = {1},
          eid = {L12},
        pages = {L12},
          doi = {10.3847/2041-8213/ade396},
archivePrefix = {arXiv},
       eprint = {2506.09572},
 primaryClass = {astro-ph.SR},
       adsurl = {https://ui.adsabs.harvard.edu/abs/2025ApJ...987L..12L}
}

@ARTICLE{Perez2025,
       author = {{Perez}, Nicolas and {Leclerc}, Armand and {Laibe}, Guillaume and {Delplace}, Pierre},
        title = "{Topology of shallow-water waves on a rotating sphere}",
      journal = {Journal of Fluid Mechanics},
         year = 2025,
        month = jan,
       volume = {1003},
          eid = {A35},
        pages = {A35},
          doi = {10.1017/jfm.2024.1228},
archivePrefix = {arXiv},
       eprint = {2404.07655},
 primaryClass = {physics.flu-dyn},
       adsurl = {https://ui.adsabs.harvard.edu/abs/2025JFM..1003A..35P}
}

@ARTICLE{Leclerc2024c,
       author = {{Leclerc}, Armand and {Laibe}, Guillaume and {Perez}, Nicolas},
        title = "{Wave topology of stellar inertial oscillations}",
      journal = {Physical Review Research},
         year = 2024,
        month = dec,
       volume = {6},
       number = {4},
          eid = {043299},
        pages = {043299},
          doi = {10.1103/PhysRevResearch.6.043299},
archivePrefix = {arXiv},
       eprint = {2411.08457},
 primaryClass = {astro-ph.SR},
       adsurl = {https://ui.adsabs.harvard.edu/abs/2024PhRvR...6d3299L}
}

@ARTICLE{Leclerc2024a,
       author = {{Leclerc}, Armand and {Jezequel}, Lucien and {Perez}, Nicolas and {Bhandare}, Asmita and {Laibe}, Guillaume and {Delplace}, Pierre},
        title = "{Exceptional ring of the buoyancy instability in stars}",
      journal = {Physical Review Research},
         year = 2024,
        month = mar,
       volume = {6},
       number = {1},
          eid = {L012055},
        pages = {L012055},
          doi = {10.1103/PhysRevResearch.6.L012055},
archivePrefix = {arXiv},
       eprint = {2311.05944},
 primaryClass = {astro-ph.SR},
       adsurl = {https://ui.adsabs.harvard.edu/abs/2024PhRvR...6a2055L}
}

@BOOK{Safronov1972,
       author = {{Safronov}, V.~S.},
        title = "{Evolution of the protoplanetary cloud and formation of the earth and planets.}",
         year = 1972,
       adsurl = {https://ui.adsabs.harvard.edu/abs/1972epcf.book.....S}
}

@INPROCEEDINGS{Lesur2023,
       author = {{Lesur}, G. and {Flock}, M. and {Ercolano}, B. and {Lin}, M. -K. and {Yang}, C. and {Barranco}, J.~A. and {Benitez-Llambay}, P. and {Goodman}, J. and {Johansen}, A. and {Klahr}, H. and {Laibe}, G. and {Lyra}, W. and {Marcus}, P.~S. and {Nelson}, R.~P. and {Squire}, J. and {Simon}, J.~B. and {Turner}, N.~J. and {Umurhan}, O.~M. and {Youdin}, A.~N.},
        title = "{Hydro-, Magnetohydro-, and Dust-Gas Dynamics of Protoplanetary Disks}",
    booktitle = {Protostars and Planets VII},
         year = 2023,
       editor = {{Inutsuka}, S. and {Aikawa}, Y. and {Muto}, T. and {Tomida}, K. and {Tamura}, M.},
       series = {Astronomical Society of the Pacific Conference Series},
       volume = {534},
        month = jul,
        pages = {465},
          doi = {10.48550/arXiv.2203.09821},
archivePrefix = {arXiv},
       eprint = {2203.09821},
 primaryClass = {astro-ph.EP},
       adsurl = {https://ui.adsabs.harvard.edu/abs/2023ASPC..534..465L}
}

@INPROCEEDINGS{Bae2023,
       author = {{Bae}, J. and {Isella}, A. and {Zhu}, Z. and {Martin}, R. and {Okuzumi}, S. and {Suriano}, S.},
        title = "{Structured Distributions of Gas and Solids in Protoplanetary Disks}",
    booktitle = {Protostars and Planets VII},
         year = 2023,
       editor = {{Inutsuka}, S. and {Aikawa}, Y. and {Muto}, T. and {Tomida}, K. and {Tamura}, M.},
       series = {Astronomical Society of the Pacific Conference Series},
       volume = {534},
        month = jul,
        pages = {423},
          doi = {10.48550/arXiv.2210.13314},
archivePrefix = {arXiv},
       eprint = {2210.13314},
 primaryClass = {astro-ph.EP},
       adsurl = {https://ui.adsabs.harvard.edu/abs/2023ASPC..534..423B}
}

@INPROCEEDINGS{Dra2023,
       author = {{Dra{\.z}kowska}, J. and {Bitsch}, B. and {Lambrechts}, M. and {Mulders}, G.~D. and {Harsono}, D. and {Vazan}, A. and {Liu}, B. and {Ormel}, C.~W. and {Kretke}, K. and {Morbidelli}, A.},
        title = "{Planet Formation Theory in the Era of ALMA and Kepler: from Pebbles to Exoplanets}",
    booktitle = {Protostars and Planets VII},
         year = 2023,
       editor = {{Inutsuka}, S. and {Aikawa}, Y. and {Muto}, T. and {Tomida}, K. and {Tamura}, M.},
       series = {Astronomical Society of the Pacific Conference Series},
       volume = {534},
        month = jul,
        pages = {717},
          doi = {10.48550/arXiv.2203.09759},
archivePrefix = {arXiv},
       eprint = {2203.09759},
 primaryClass = {astro-ph.EP},
       adsurl = {https://ui.adsabs.harvard.edu/abs/2023ASPC..534..717D}
}

@ARTICLE{Teague2025,
       author = {{Teague}, Richard and {Benisty}, Myriam and {Facchini}, Stefano and {Fukagawa}, Misato and {Pinte}, Christophe and {Andrews}, Sean M. and {Bae}, Jaehan and {Barraza-Alfaro}, Marcelo and {Cataldi}, Gianni and {Cuello}, Nicol{\'a}s and {Curone}, Pietro and {Czekala}, Ian and {Fasano}, Daniele and {Flock}, Mario and {Galloway-Sprietsma}, Maria and {Garg}, Himanshi and {Hall}, Cassandra and {Hammond}, Iain and {Hilder}, Thomas and {Huang}, Jane and {Ilee}, John D. and {Izquierdo}, Andr{\'e}s F. and {Kanagawa}, Kazuhiro and {Lesur}, Geoffroy and {Lodato}, Giuseppe and {Longarini}, Cristiano and {Loomis}, Ryan A. and {Masset}, Fr{\'e}d{\'e}ric and {Menard}, Francois and {Orihara}, Ryuta and {Price}, Daniel J. and {Rosotti}, Giovanni and {Stadler}, Jochen and {Testi}, Leonardo and {Yen}, Hsi-Wei and {Wafflard-Fernandez}, Gaylor and {Wilner}, David J. and {Winter}, Andrew J. and {W{\"o}lfer}, Lisa and {Yoshida}, Tomohiro C. and {Zawadzki}, Brianna},
        title = "{exoALMA. I. Science Goals, Project Design, and Data Products}",
      journal = {\apjl},
         year = 2025,
        month = may,
       volume = {984},
       number = {1},
          eid = {L6},
        pages = {L6},
          doi = {10.3847/2041-8213/adc43b},
archivePrefix = {arXiv},
       eprint = {2504.18688},
 primaryClass = {astro-ph.EP},
       adsurl = {https://ui.adsabs.harvard.edu/abs/2025ApJ...984L...6T}
}

@INPROCEEDINGS{Pinte2023,
       author = {{Pinte}, C. and {Teague}, R. and {Flaherty}, K. and {Hall}, C. and {Facchini}, S. and {Casassus}, S.},
        title = "{Kinematic Structures in Planet-Forming Disks}",
    booktitle = {Protostars and Planets VII},
         year = 2023,
       editor = {{Inutsuka}, S. and {Aikawa}, Y. and {Muto}, T. and {Tomida}, K. and {Tamura}, M.},
       series = {Astronomical Society of the Pacific Conference Series},
       volume = {534},
        month = jul,
        pages = {645},
          doi = {10.48550/arXiv.2203.09528},
archivePrefix = {arXiv},
       eprint = {2203.09528},
 primaryClass = {astro-ph.EP},
       adsurl = {https://ui.adsabs.harvard.edu/abs/2023ASPC..534..645P}
}

@ARTICLE{Solheim1998,
       author = {{Solheim}, J. -E. and {Provencal}, J.~L. and {Bradley}, P.~A. and {Vauclair}, G. and {Barstow}, M.~A. and {Kepler}, S.~O. and {Fontaine}, G. and {Grauer}, A.~D. and {Winget}, D.~E. and {Marar}, T.~M.~K. and {Leibowitz}, E.~M. and {Emanuelsen}, P. -I. and {Chevreton}, M. and {Dolez}, N. and {Kanaan}, A. and {Bergeron}, P. and {Claver}, C.~F. and {Clemens}, J.~C. and {Kleinman}, S.~J. and {Hine}, B.~P. and {Seetha}, S. and {Ashoka}, B.~N. and {Mazeh}, T. and {Sansom}, A.~E. and {Tweedy}, R.~W. and {Meistas}, E.~G. and {Bruvold}, A. and {Massacand}, C.~M.},
        title = "{Whole Earth Telescope observations of AM Canum Venaticorum - discoseismology at last}",
      journal = {\aap},
         year = 1998,
        month = apr,
       volume = {332},
        pages = {939-957},
       adsurl = {https://ui.adsabs.harvard.edu/abs/1998A&A...332..939S}
}

@ARTICLE{Tsang2009,
       author = {{Tsang}, David and {Lai}, Dong},
        title = "{Corotational damping of discoseismic c modes in black hole accretion discs}",
      journal = {\mnras},
         year = 2009,
        month = mar,
       volume = {393},
       number = {3},
        pages = {992-998},
          doi = {10.1111/j.1365-2966.2008.14228.x},
archivePrefix = {arXiv},
       eprint = {0810.1299},
 primaryClass = {astro-ph},
       adsurl = {https://ui.adsabs.harvard.edu/abs/2009MNRAS.393..992T}
}

@ARTICLE{Tsang2013,
       author = {{Tsang}, David and {Butsky}, Iryna},
        title = "{Iron line variability of discoseismic corrugation modes}",
      journal = {\mnras},
         year = 2013,
        month = oct,
       volume = {435},
       number = {1},
        pages = {749-765},
          doi = {10.1093/mnras/stt1334},
archivePrefix = {arXiv},
       eprint = {1307.4971},
 primaryClass = {astro-ph.HE},
       adsurl = {https://ui.adsabs.harvard.edu/abs/2013MNRAS.435..749T}
}

@ARTICLE{OR2020,
       author = {{Ortega-Rodr{\'\i}guez}, M. and {Sol{\'\i}s-S{\'a}nchez}, H. and {{\'A}lvarez-Garc{\'\i}a}, L. and {Dodero-Rojas}, E.},
        title = "{On twin peak quasi-periodic oscillations resulting from the interaction between discoseismic modes and turbulence in accretion discs around black holes}",
      journal = {\mnras},
         year = 2020,
        month = feb,
       volume = {492},
       number = {2},
        pages = {1755-1760},
          doi = {10.1093/mnras/stz3541},
archivePrefix = {arXiv},
       eprint = {1912.09527},
 primaryClass = {astro-ph.HE},
       adsurl = {https://ui.adsabs.harvard.edu/abs/2020MNRAS.492.1755O}
}

@ARTICLE{Dewberry2018,
       author = {{Dewberry}, Janosz W. and {Latter}, Henrik N. and {Ogilvie}, Gordon I.},
        title = "{Quasi-periodic oscillations and the global modes of relativistic, MHD accretion discs}",
      journal = {\mnras},
         year = 2018,
        month = may,
       volume = {476},
       number = {3},
        pages = {4085-4103},
          doi = {10.1093/mnras/sty385},
archivePrefix = {arXiv},
       eprint = {1802.04792},
 primaryClass = {astro-ph.HE},
       adsurl = {https://ui.adsabs.harvard.edu/abs/2018MNRAS.476.4085D}
}

@ARTICLE{Dewberry2020a,
       author = {{Dewberry}, Janosz W. and {Latter}, Henrik N. and {Ogilvie}, Gordon I. and {Fromang}, Sebastien},
        title = "{HFQPOs and discoseismic mode excitation in eccentric, relativistic discs. I. Hydrodynamic simulations}",
      journal = {\mnras},
         year = 2020,
        month = sep,
       volume = {497},
       number = {1},
        pages = {435-450},
          doi = {10.1093/mnras/staa1897},
archivePrefix = {arXiv},
       eprint = {1910.10696},
 primaryClass = {astro-ph.HE},
       adsurl = {https://ui.adsabs.harvard.edu/abs/2020MNRAS.497..435D}
}

@ARTICLE{Dewberry2020b,
       author = {{Dewberry}, Janosz W. and {Latter}, Henrik N. and {Ogilvie}, Gordon I. and {Fromang}, Sebastien},
        title = "{HFQPOs and discoseismic mode excitation in eccentric, relativistic discs. II. Magnetohydrodynamic simulations}",
      journal = {\mnras},
         year = 2020,
        month = sep,
       volume = {497},
       number = {1},
        pages = {451-465},
          doi = {10.1093/mnras/staa1898},
archivePrefix = {arXiv},
       eprint = {2006.16266},
 primaryClass = {astro-ph.HE},
       adsurl = {https://ui.adsabs.harvard.edu/abs/2020MNRAS.497..451D}
}

@ARTICLE{Kato2024,
       author = {{Kato}, Shoji},
        title = "{Damping of disco-seismic C-mode oscillations at the sonic radius of discs}",
      journal = {\mnras},
         year = 2024,
        month = feb,
       volume = {528},
       number = {2},
        pages = {1408-1421},
          doi = {10.1093/mnras/stae027},
       adsurl = {https://ui.adsabs.harvard.edu/abs/2024MNRAS.528.1408K}
}

@article{marley1991nonradial,
  title={Nonradial oscillations of Saturn},
  author={Marley, Mark S},
  journal={Icarus},
  volume={94},
  number={2},
  pages={420--435},
  year={1991},
  publisher={Elsevier}
}

@article{fuller2014saturn,
  title={Saturn ring seismology: Evidence for stable stratification in the deep interior of Saturn},
  author={Fuller, Jim},
  journal={icarus},
  volume={242},
  pages={283--296},
  year={2014},
  publisher={Elsevier}
}

@article{lovelace1999rossby,
  title={Rossby wave instability of Keplerian accretion disks},
  author={Lovelace, RVE and Li, H and Colgate, SA and Nelson, AF},
  journal={The Astrophysical Journal},
  volume={513},
  number={2},
  pages={805},
  year={1999},
  publisher={IOP Publishing}
}

@BOOK{Aerts2010,
       author = {{Aerts}, Conny and {Christensen-Dalsgaard}, J{\o}rgen and {Kurtz}, Donald W.},
        title = "{Asteroseismology}",
         year = 2010,
          doi = {10.1007/978-1-4020-5803-5},
       adsurl = {https://ui.adsabs.harvard.edu/abs/2010aste.book.....A}
}

@ARTICLE{Zhuravlev2019,
       author = {{Zhuravlev}, V.~V.},
        title = "{On the nature of the resonant drag instability of dust streaming in protoplanetary disc}",
      journal = {\mnras},
         year = 2019,
        month = nov,
       volume = {489},
       number = {3},
        pages = {3850-3869},
          doi = {10.1093/mnras/stz2390},
archivePrefix = {arXiv},
       eprint = {1907.09626},
 primaryClass = {astro-ph.EP},
       adsurl = {https://ui.adsabs.harvard.edu/abs/2019MNRAS.489.3850Z}
}

@ARTICLE{Lehmann2023,
       author = {{Lehmann}, Marius and {Lin}, Min-Kai},
        title = "{Instabilities in dusty non-isothermal protoplanetary discs}",
      journal = {\mnras},
         year = 2023,
        month = jul,
       volume = {522},
       number = {4},
        pages = {5892-5930},
          doi = {10.1093/mnras/stad1349},
archivePrefix = {arXiv},
       eprint = {2305.02362},
 primaryClass = {astro-ph.EP},
       adsurl = {https://ui.adsabs.harvard.edu/abs/2023MNRAS.522.5892L}
}

@ARTICLE{Paardekooper2025,
       author = {{Paardekooper}, Sijme-Jan and {Aly}, Hossam},
        title = "{Resonant drag instabilities for polydisperse dust: II. The streaming and settling instabilities}",
      journal = {\aap},
         year = 2025,
        month = may,
       volume = {697},
          eid = {A40},
        pages = {A40},
          doi = {10.1051/0004-6361/202453496},
archivePrefix = {arXiv},
       eprint = {2503.09265},
 primaryClass = {astro-ph.EP},
       adsurl = {https://ui.adsabs.harvard.edu/abs/2025A&A...697A..40P}
}

@article{okazaki1987,
  title={Global trapped oscillations of relativistic accretion disks},
  author={Okazaki, Atsuo T and Kato, Shoji and Fukue, Jun},
  journal={Publications of the Astronomical Society of Japan},
  volume={39},
  number={3},
  pages={457--473},
  year={1987},
  publisher={Oxford University Press}
}

@article{iga2001,
  title={Transition modes in stratified compressible fluids},
  author={Iga, Keita},
  journal={Fluid dynamics research},
  volume={28},
  number={6},
  pages={465},
  year={2001},
  publisher={IOP Publishing}
}

\appendix
\section{Constant of motion}
The evolution equation of the perturbations $X(r,k_z,t) $ is
\begin{equation}
    i\partial_t X= \mathcal{H}X.
\end{equation}
As $\mathcal{H}$ is self-adjoint with respect to the inner product 
\begin{equation}
    \langle X,{Y}\rangle = \int \dd r\; X^{\top*}\cdot Y,
\end{equation}
where $\cdot$ is the matricial product, such that $\langle X,\mathcal{H}{Y}\rangle = \langle \mathcal{H}X,{Y}\rangle$ for boundary conditions with zero radial velocity. One then has
\begin{eqnarray}
    i\partial_t \langle X,X\rangle &=& \langle -i\partial_tX,X\rangle + \langle X,i\partial_tX\rangle,\\
    &=& -\langle \mathcal{H}X,X\rangle + \langle X,\mathcal{H}X\rangle,\\
    &=& -\langle X,\mathcal{H}X\rangle + \langle X,\mathcal{H}X\rangle,\\
    &=& 0.
\end{eqnarray}
$I = \frac{1}{2}\langle X,X\rangle$ is then a constant of motion, given by the self-adjointess of the evolution operator $\mathcal{H}$. Equation~\eqref{eq:constantOfMotion} gives it in dimensional form, where one reads that it is a pseudo-energy of the wave, as the sum of kinetic and pressure energy contributions.





\section{Self-adjointness of \texorpdfstring{$\mathcal{H}$}{}}
We show here that $\mathcal{H}$ is self-adjoint on the Hilbert space $\{X = \begin{pmatrix}u_z & u_r & u_\theta & h \end{pmatrix}^\top, u_r(r_0)=u_r(r_1)=0\}$ with respect to the canonical inner product $\langle X_1,X_2\rangle = \int \dd r X_1^* X_2$.

\begin{eqnarray}
    \langle X_1,\mathcal{H}X_2\rangle &=& \int\dd r\; \bigg( c_\mathrm{s}k_z(u_{z,1}^* h_2 + h_1^*u_{z,2})+i\kappa(u_{\theta,1}^*u_{r,2}-u_{r,1}^*u_{\theta,2})\bigg) \\
    &+& \int \dd r \; u_{r,1}^* \left(ic_\mathrm{s}\partial_rh_2 +i\frac{c_\mathrm{s}^\prime}{2}h_2 -iSh_2\right) + \int \dd r \; h_{1}^* \left(ic_\mathrm{s}\partial_ru_{r,2} +i\frac{c_\mathrm{s}^\prime}{2}u_{r,2} +iSu_{r,2}\right),\nonumber\\
    &=& \int\dd r\; \bigg( c_\mathrm{s}k_z(u_{z,1}^* h_2 + h_1^*u_{z,2})+i\kappa(u_{\theta,1}^*u_{r,2}-u_{r,1}^*u_{\theta,2})\bigg) \\
    &+& [ic_\mathrm{s}u_{r,1}^* h_2] - \int \dd r \left(ic_\mathrm{s}\partial_ru_{r,1}^* +i\frac{c_\mathrm{s}^\prime}{2}u_{r,1}^* +iSu_{r,1}^*\right) h_2 + [ic_\mathrm{s} h_{1}^* u_{r,2}] - \int \dd r \left(ic_\mathrm{s}\partial_rh_{1}^* +i\frac{c_\mathrm{s}^\prime}{2}h_{1}^* -iSh_{1}^*\right)u_{r,2},\nonumber\\
    &=& \int\dd r\; \bigg( (c_\mathrm{s}k_z u_{z,1})^* h_2 + (c_\mathrm{s}k_zh_1)^*u_{z,2}+(-i\kappa u_{\theta,1})^*u_{r,2}+(i\kappa u_{r,1})^*u_{\theta,2})\bigg) \\ 
    &+& [ic_\mathrm{s}u_{r,1}^* h_2] + \int \dd r \left(ic_\mathrm{s}\partial_ru_{r,1} +i\frac{c_\mathrm{s}^\prime}{2}u_{r,1} +iSu_{r,1}\right)^* h_2+ [ic_\mathrm{s} h_{1}^* u_{r,2}] + \int \dd r \left(ic_\mathrm{s}\partial_rh_{1} +i\frac{c_\mathrm{s}^\prime}{2}h_{1} -iSh_{1}\right)^*u_{r,2},\nonumber\\
    &=& \langle \mathcal{H}X_1,X_2\rangle + [ic_\mathrm{s}(u_{r,1}^* h_2+h_{1}^* u_{r,2})].
\end{eqnarray}
The bracket term goes to zero for impenetrable boundaries. $\mathcal{H}$ is therefore self-adjoint.

\newpage 
\section{Profiles of modes}
\label{app:profiles_modes}

\begin{figure}[h!]
    \centering
    \includegraphics[width=\linewidth]{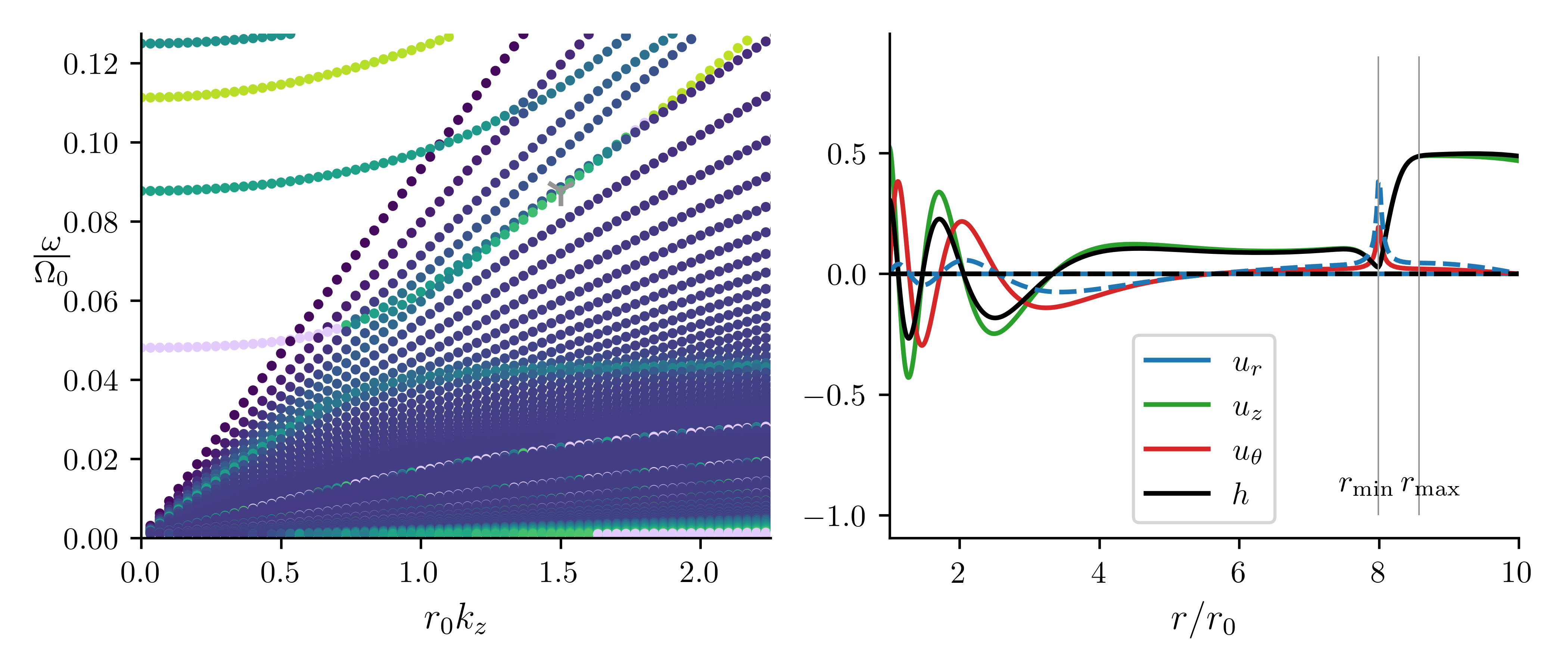}
    \includegraphics[width=\linewidth]{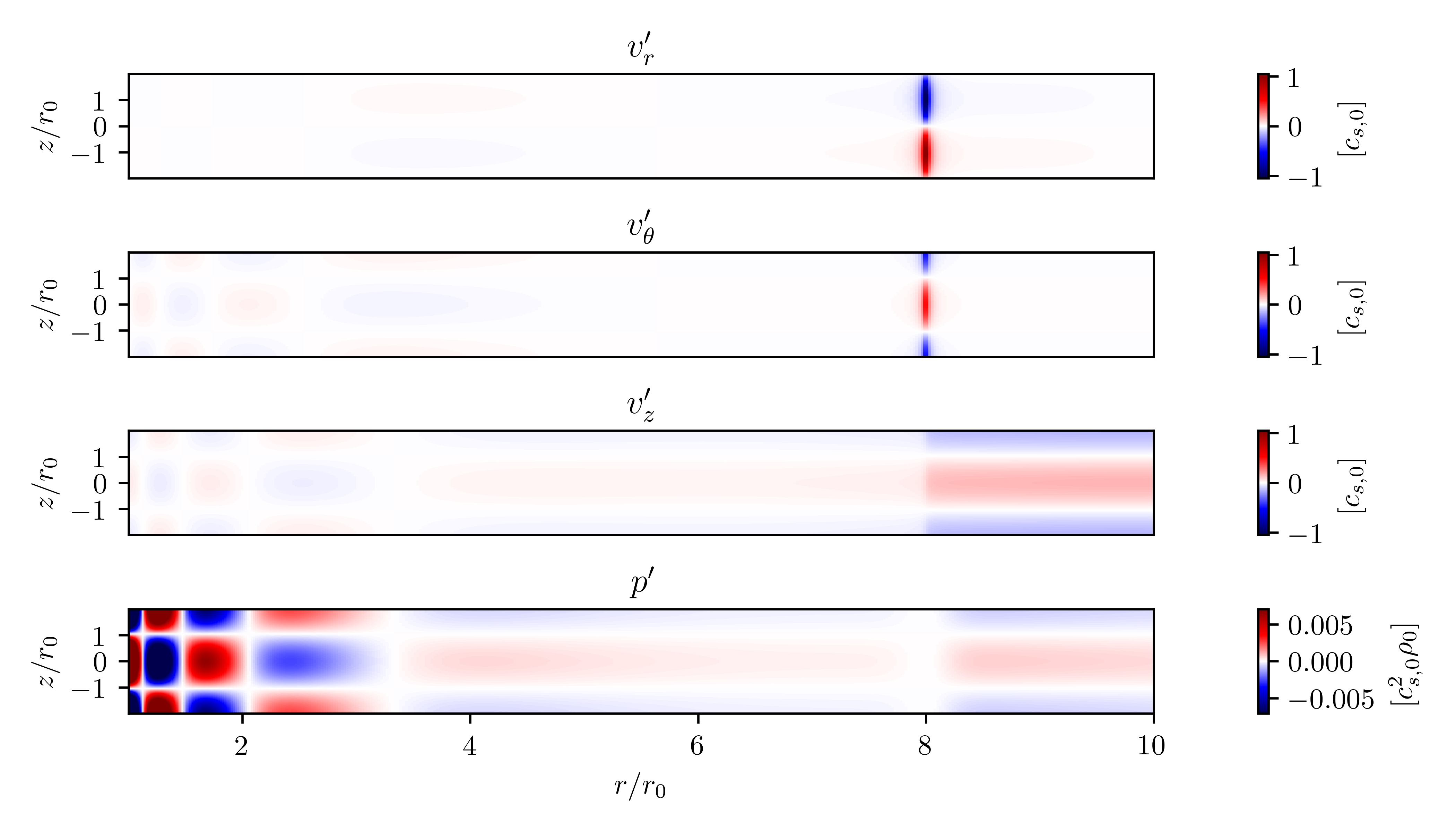}
    \caption{Relative amplitudes of the components of a fundamental $p$-mode with large $k_{z}$, of topological origin, induced by a pressure bump located close to $r_\mathrm{max}$. The amplitudes are displayed as functions of the distance to the central object in the mid-plane (top right panel) and for a $\left( r,z\right)$ slice of the disc (bottom panel). The selected mode is marked with a cross in the top left panel (same as Fig.~\ref{fig:monotonic}). This mode {color{red}haveamplitudes(?)} in $\left(u_{z},h \right)$ and as such, is of acoustic nature. Interrestingly, it mixes with both the fundamental $r$-mode of the gap and an inertial mode of the inner disc, which both have amplitudes in $u_{r}$ and $u_{\theta}$. It is more spatially extended around $r_\mathrm{max}$ than the fundamental $r$-mode is around $r_\mathrm{min}$ (see below).}
    \label{fig:eigenprofiles_acousticMode_gap}
\end{figure}

\begin{figure}[h!]
    \centering
    \includegraphics[width=\linewidth]{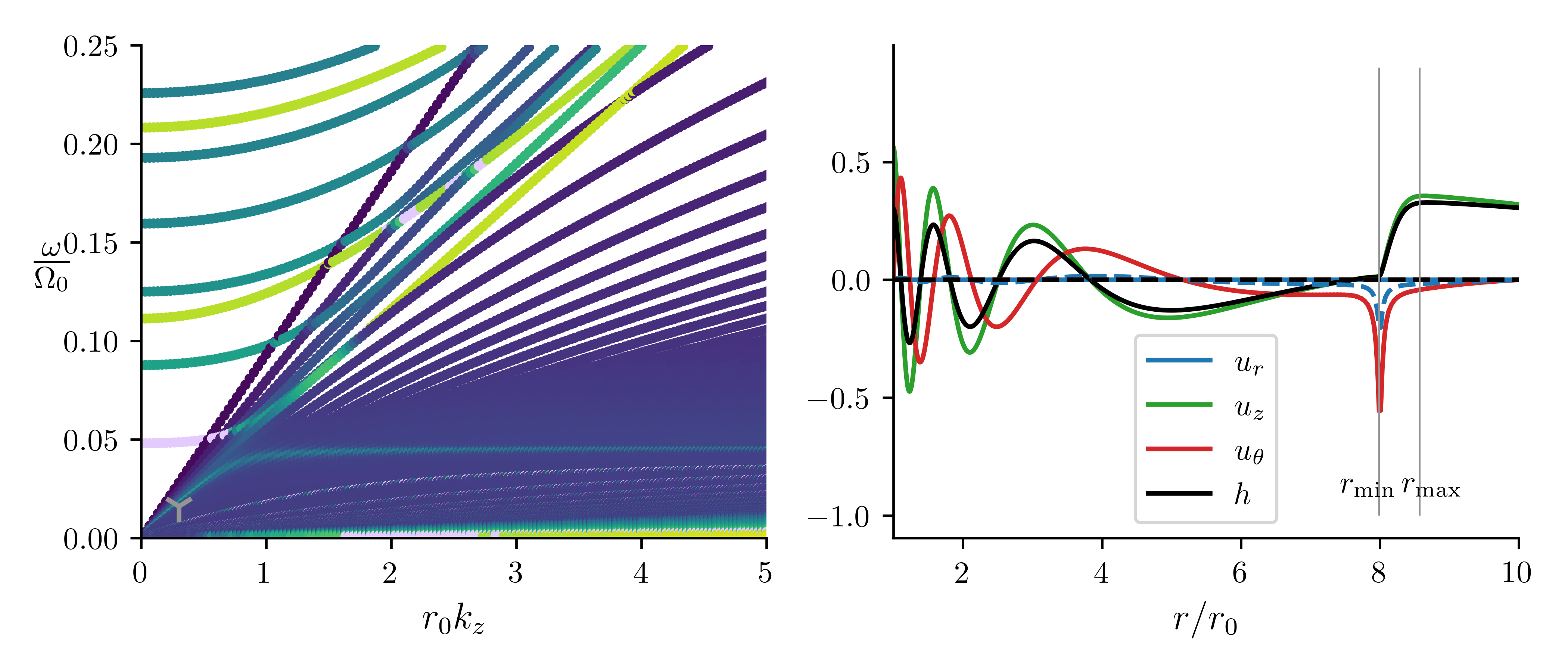}
    \includegraphics[width=\linewidth]{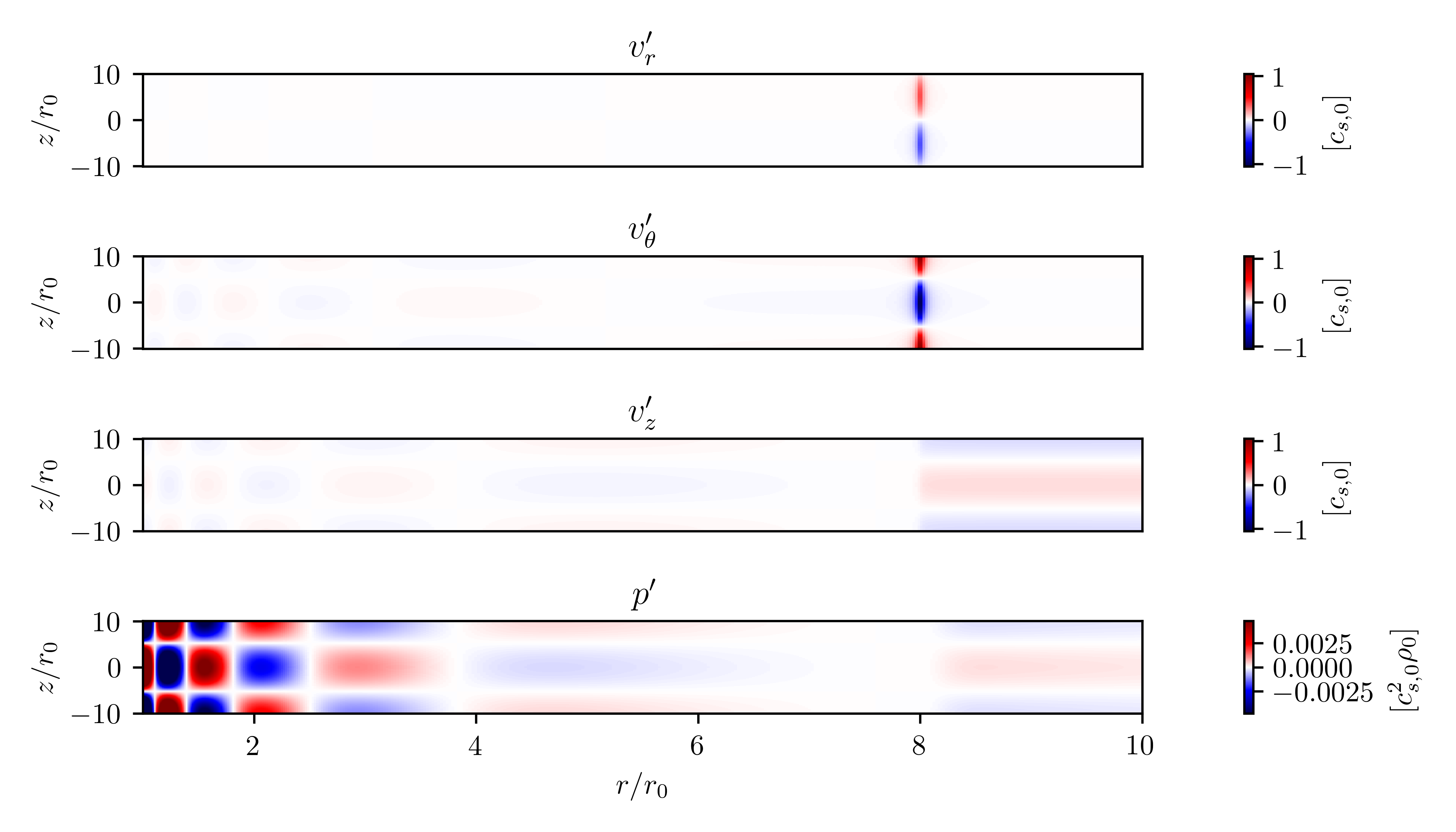}
    \caption{Same as Fig.~\ref{fig:eigenprofiles_acousticMode_gap}, but for a mode a low $k_z$ (note the different $z$ scale). The properties of this mode are similar to the one displayed in Fig.~\ref{fig:eigenprofiles_acousticMode_gap}. }
\label{fig:eigenprofiles_acousticMode_gap_lowK}
\end{figure}

\begin{figure}[h!]
    \centering
    \includegraphics[width=\linewidth]{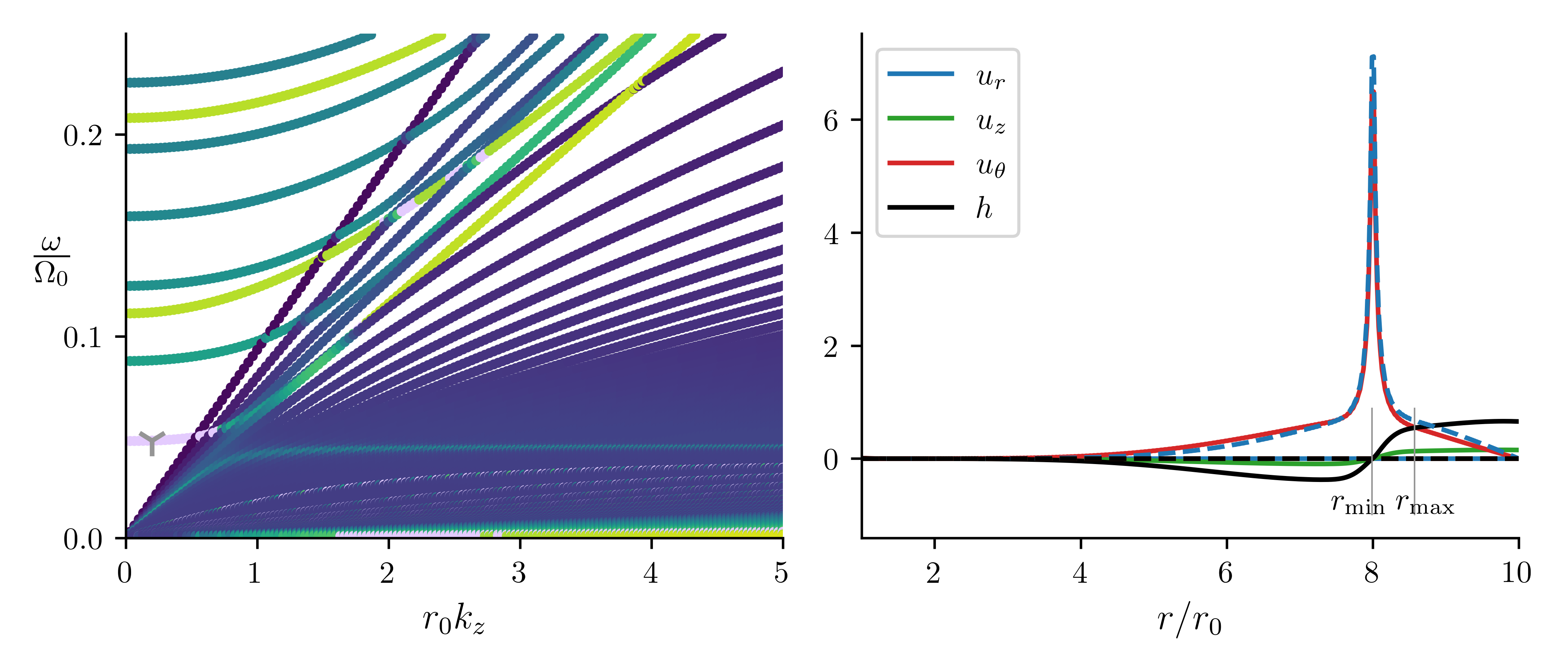}
    \includegraphics[width=\linewidth]{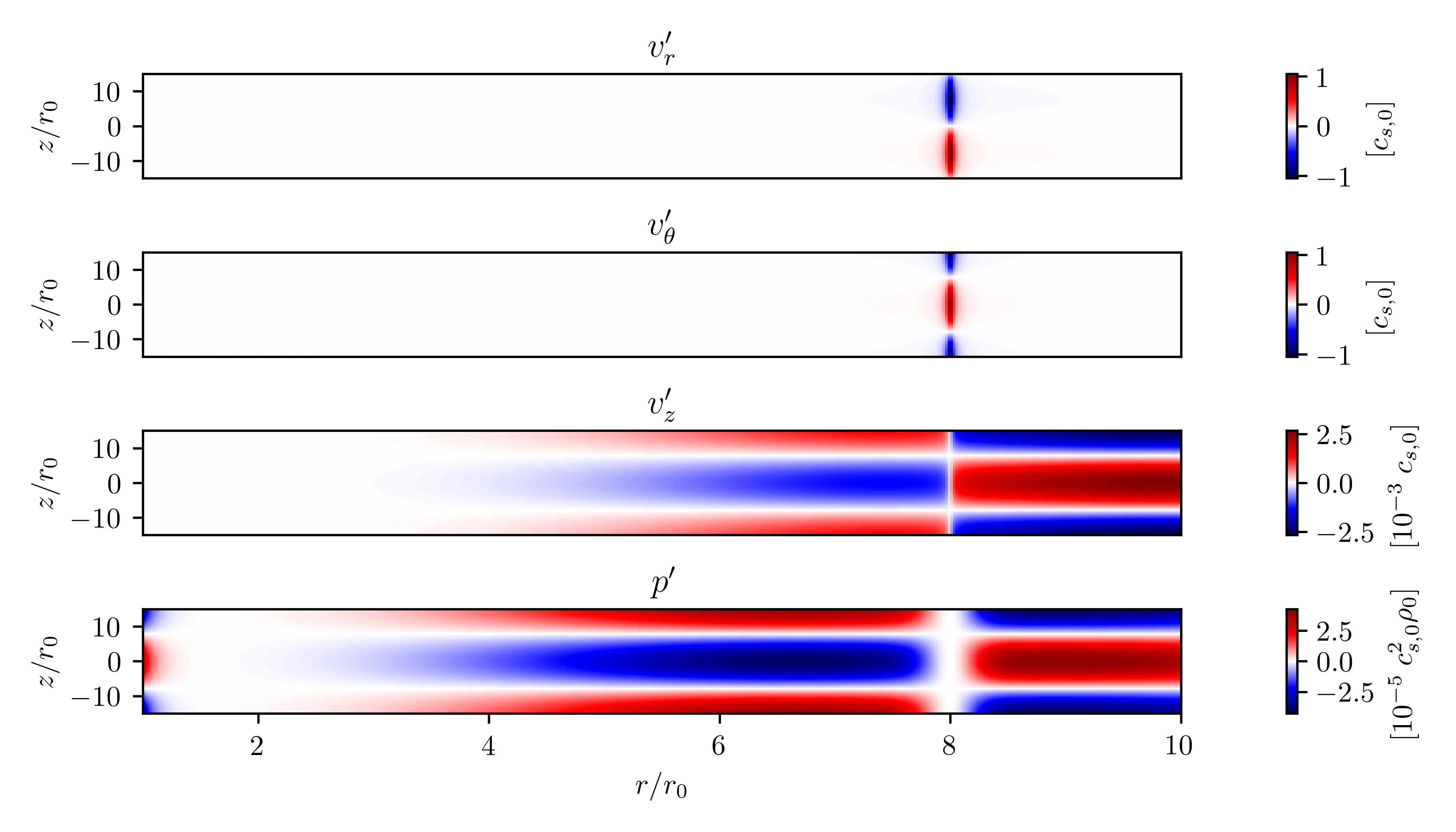}
    \caption{Same as Fig.~\ref{fig:eigenprofiles_acousticMode_gap} for the the fundamental $r$-mode, topologically induced by the existence of the a pressure gap. The mode exhibits strong horizontal motions around $r_{\rm min}$ (note the scales colorbars), where the amplitude peaks. The mode extends outwards of the gap, and has non-zero amplitude at $r_{\rm max}$, where grains may eventually pile-up. The mode mixes with acoustic vibrations outside of $r_{\rm min}$.}
    \label{fig:eigenprofiles_gap_lowK}
\end{figure}

\begin{figure}[h!]
    \centering
    \includegraphics[width=\linewidth]{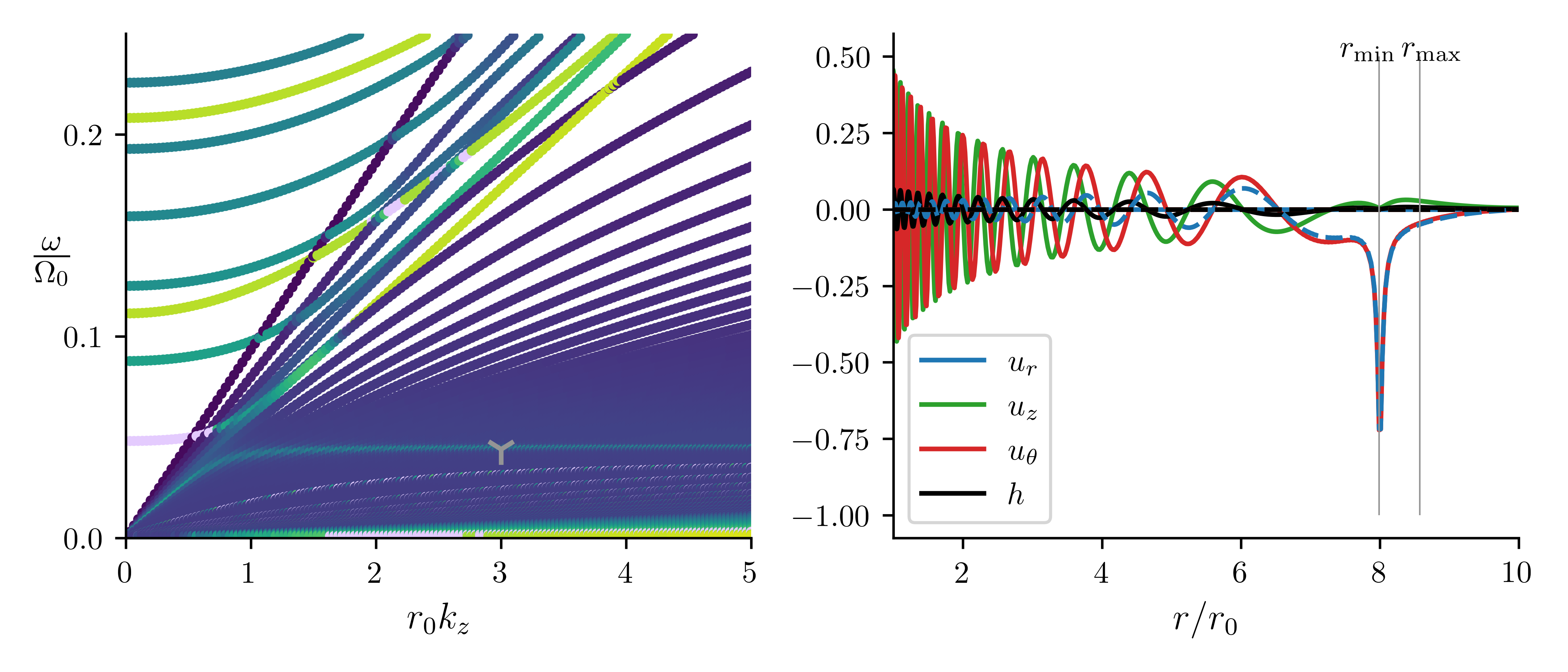}
    \includegraphics[width=\linewidth]{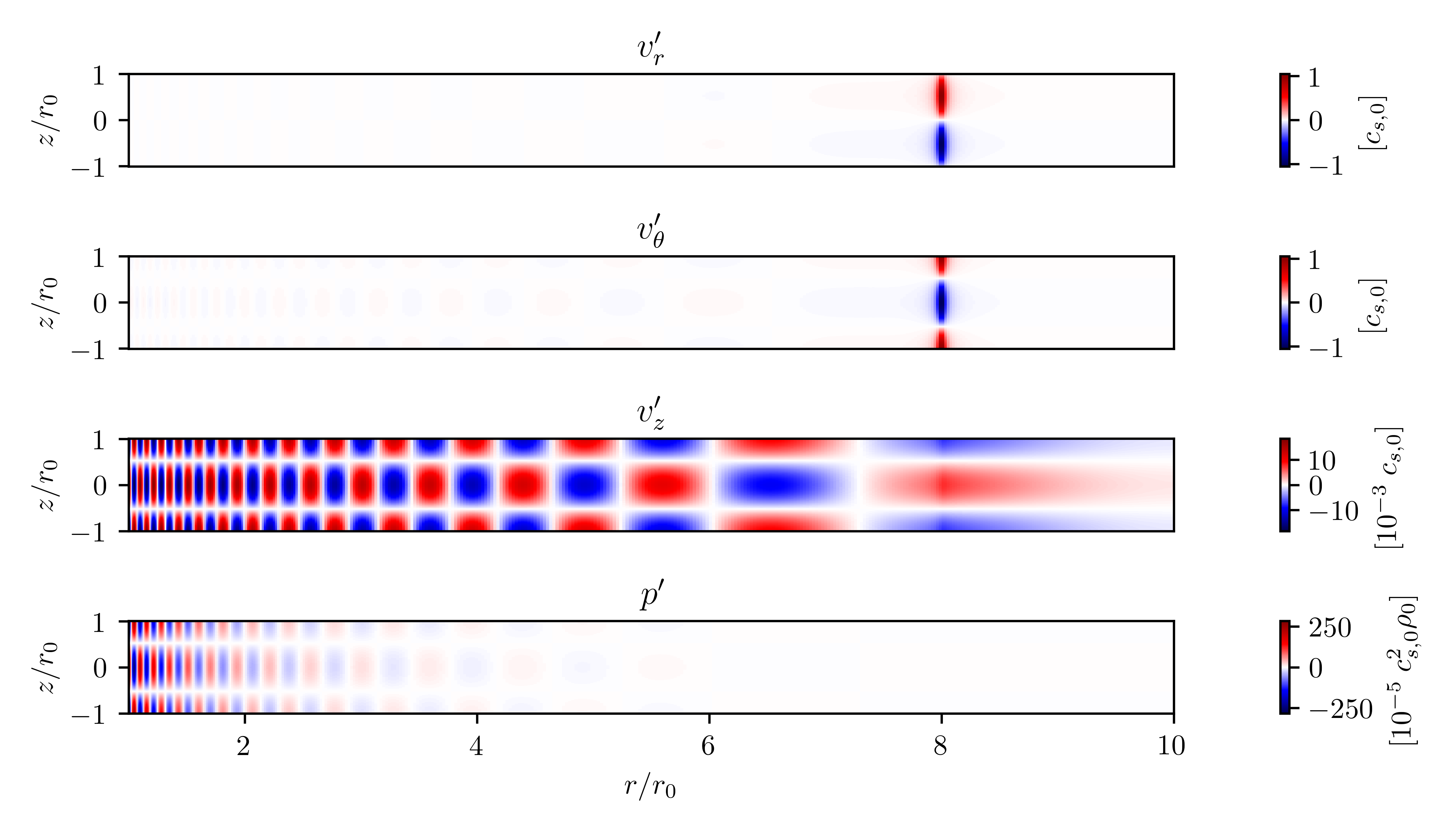}
    \caption{Same as Fig.~\ref{fig:eigenprofiles_gap_lowK}, but for larger values of $k_z$ (note the different scale on $z$). The fundamental vibration of the gap couples with an inertial vibration of the rest of the disc, such that vertical motions have comparable amplitudes with horizontal motions within the gap (note the scales colorbars).}
    \label{fig:eigenprofiles_gap_highK}
\end{figure}

\newpage
\begin{figure}[h!]
    \centering
    \includegraphics[width=0.8\linewidth]{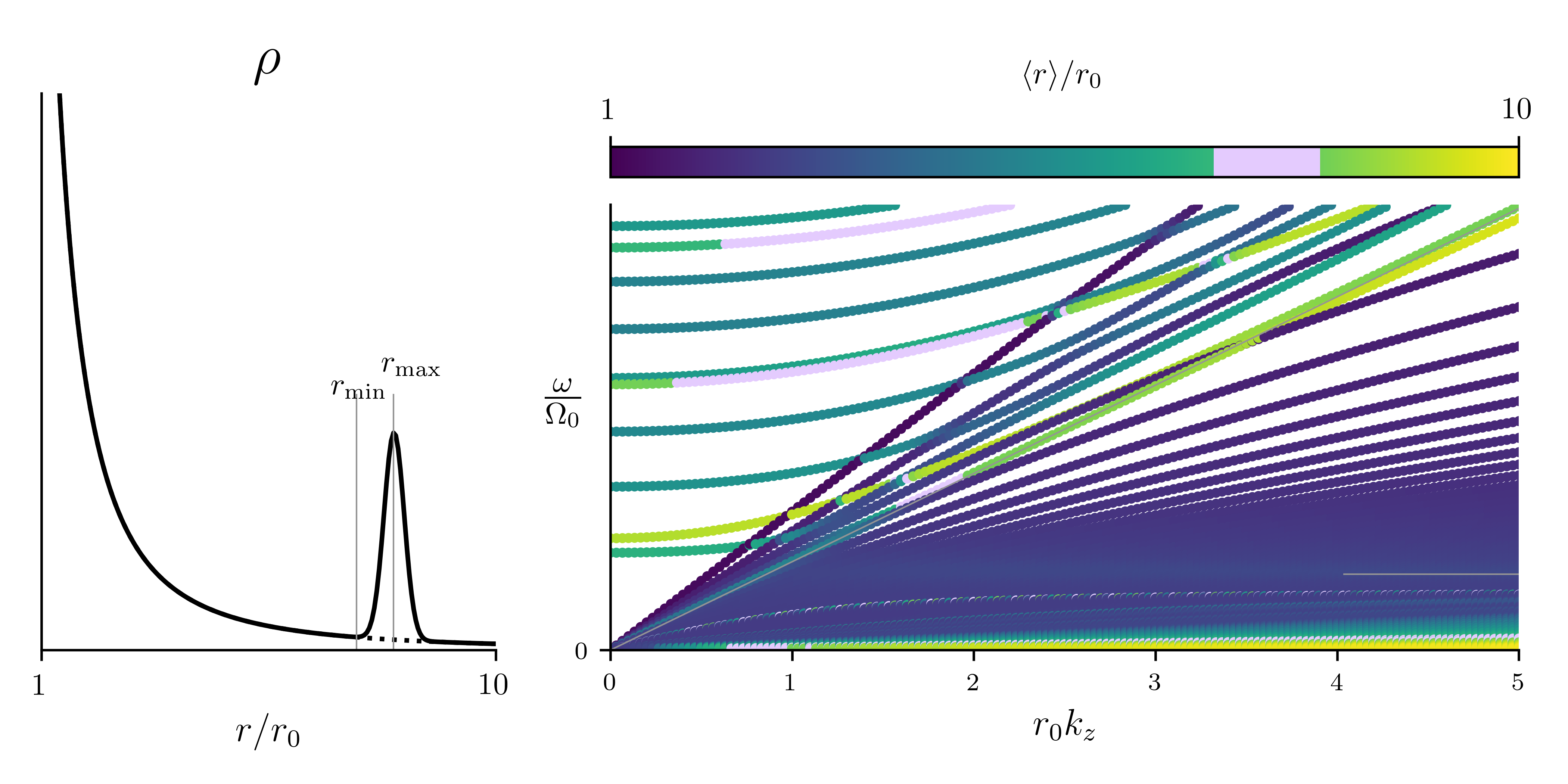}
    \includegraphics[width=0.9\linewidth]{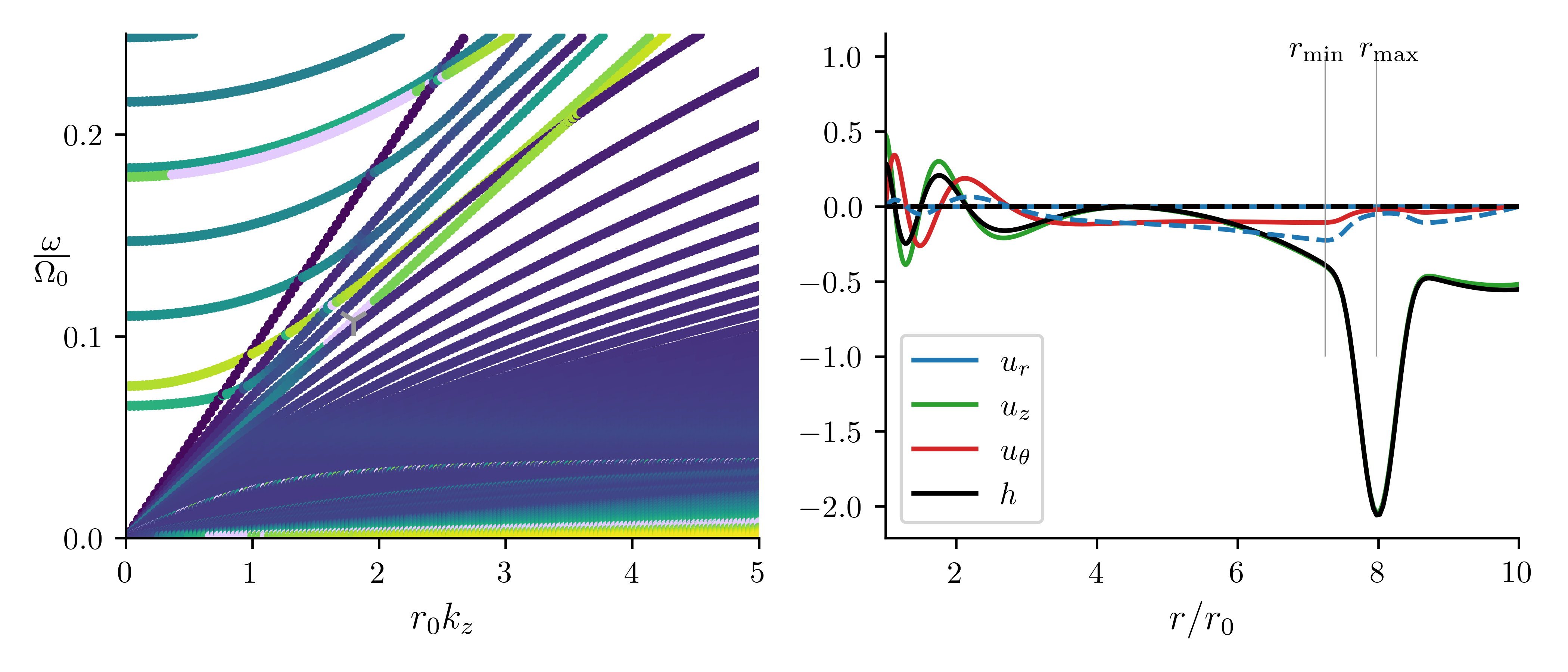}
    \includegraphics[width=0.9\linewidth]{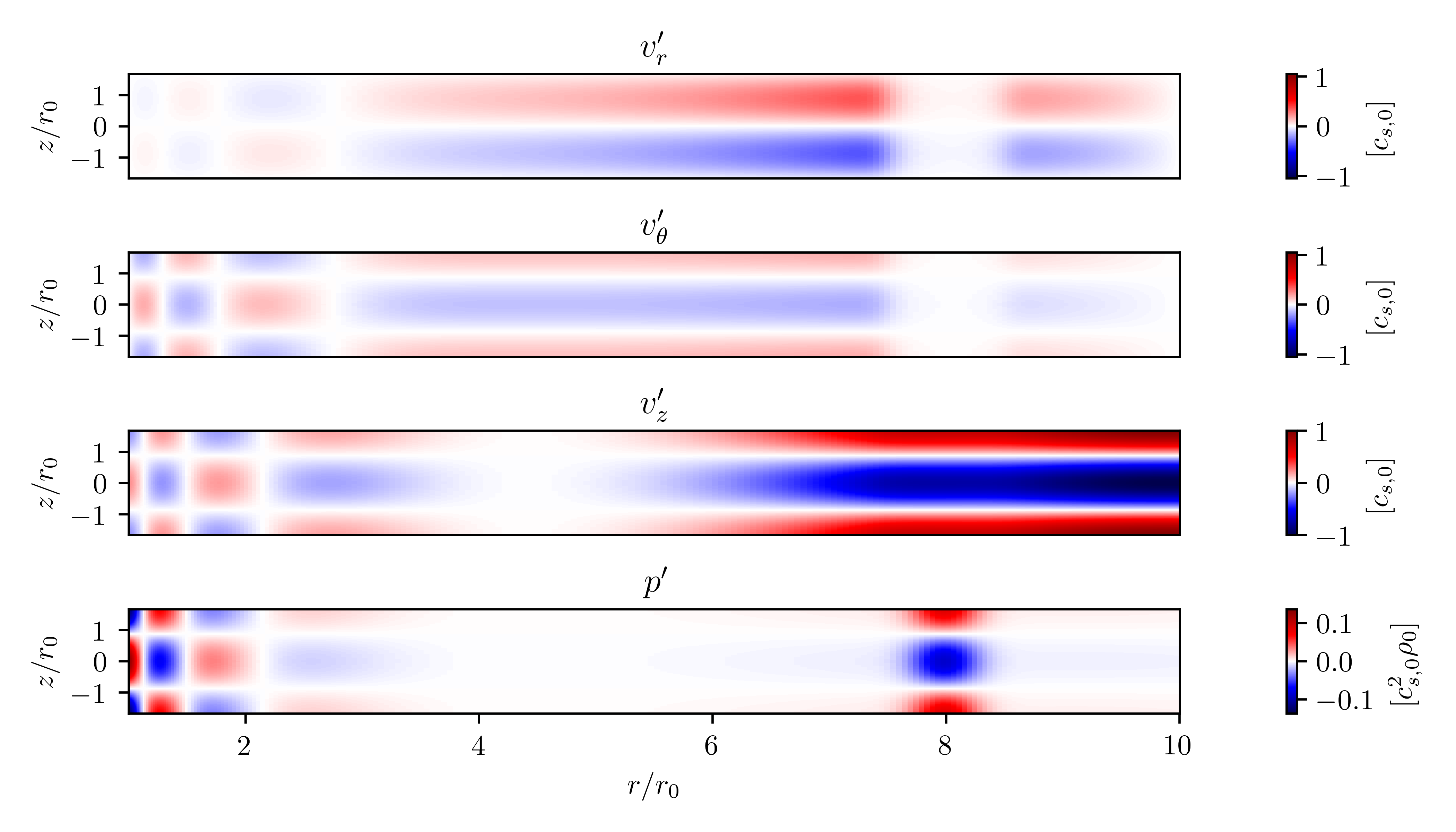}
    \caption{Same as Fig.~\ref{fig:eigenprofiles_acousticMode_gap}, but for a disc having a pressure bump. The parameters of the bump are similar to the ones used to parametrize the gap, with a reversed amplitude $A = -20$. The mode is essentially of acoustic nature, while it mixes with inertial modes in the inner regions. The inertial topological mode associated to the presence of a pressure minimum at the inner edge is present, but difficult to distinguish since it is very delocalized.}
    \label{fig:eigenprofiles_bump}
\end{figure}

\end{document}